\documentclass[reprint,aps,prl,superscriptaddress,floatfix,twocolumn,nofootinbib]{revtex4-2}

\usepackage[utf8]{inputenc}
\usepackage[T1]{fontenc}
\usepackage{amsmath,amssymb,amsfonts}
\usepackage{bm}
\usepackage{graphicx}
\usepackage{xcolor}
\usepackage{float}      
\usepackage{makecell}   

\usepackage{hyperref}
\hypersetup{colorlinks=true,linkcolor=blue,citecolor=blue,urlcolor=blue}

\newcommand{\bk}{\mathbf{k}}
\newcommand{\bq}{\mathbf{q}}

\newcommand{\bA}{\mathbf{A}}

\newcommand{\eps}{\varepsilon}
\newcommand{\dd}{\mathrm{d}}
\newcommand{\ii}{\mathrm{i}}

\newcommand{\wt}{\widetilde}
\newcommand{\Tr}{\operatorname{Tr}}
\newcommand{\order}[1]{\mathcal{O}\!\left(#1\right)}

\newcommand{\calQ}{\mathcal{Q}}

\newcommand{\calB}{\mathcal{B}}
\newcommand{\calC}{\mathcal{C}}
\newcommand{\calD}{\mathcal{D}}

\begin{document}

\title{Three Quantum-Geometric Contributions to Cubic Orbital Magnetization}

\author{Tohid Farajollahpour}
\email{tohidfrjpr@gmail.com}
\affiliation{Department of Physics, Norwegian University of Science and Technology (NTNU), NO-7491 Trondheim, Norway}
\affiliation{Department of Physics, Brock University, St. Catharines, Ontario L2S 3A1, Canada}

\begin{abstract}
In noncentrosymmetric metals such as $C_{3v}$ topological-insulator surfaces, moir\'e heterobilayers, and zincblende crystals, point-group symmetry can forbid the linear and quadratic electric-field-induced orbital magnetization, leaving the cubic response as the leading signal. Using a Ward-complete finite-momentum cubic Kubo kernel with an antisymmetric linear-in-$q$ projection, we show that the dc response separates into three quantum-geometric channels. These are a mixed electric-magnetic positional-shift quadrupole, a quantum-metric drift term, and an orbital-moment octupole. The three contributions share the same point-group symmetry but differ in their lifetime, frequency, and gate fingerprints. For a warped $C_{3v}$ surface the metric channel obeys the cutoff-independent law $\bar{\chi}_G \propto \mu^{-2}$. We propose third-harmonic magneto-optical Kerr spectroscopy as an experimental route.
\end{abstract}

\maketitle

{\color{blue}\textit{Introduction}.}\enspace
Quantum geometry of Bloch states has become a unifying language for nonlinear responses in noncentrosymmetric crystals~\cite{liu2024quantum,verma2026hidden,jiang2025revealing,nagaosa2024nonreciprocal}. At second order, the Berry-curvature dipole produces the nonlinear Hall effect~\cite{sodemann2015quantum,ma2019observation}, while the quantum metric gives a distinct intrinsic channel recently observed in topological antiferromagnets~\cite{wang2023quantum,gao2023quantum}. At third order, quantum-geometric multipoles such as the quantum-metric quadrupole~\cite{liu2025giant,yu2025quantum}, Berry-curvature quadrupole~\cite{sankar2024experimental,li2024quantum,Tohid2025a,Tohid2025b}, and Berry-connection polarizability~\cite{lai2021third,liu2022berry} generate cubic Hall and longitudinal currents~\cite{zhang2023higher,mandal2024quantum,fang2024quantum,chu2025third,das2023intrinsic,he2024third}. Much less is known for nonlinear orbital magnetization. Recently, Qiang \emph{et al.} showed that a quadratic electric-field-induced orbital magnetization is governed by a quantum Christoffel symbol~\cite{qiang2026quantum}, and Qian \emph{et al.} detected this response by second-harmonic magneto-optical Kerr spectroscopy in WTe$_2$~\cite{qian2026probing}. In several central noncentrosymmetric platforms, however, symmetry removes the quadratic magnetization channel. This occurs on $C_{3v}$ topological-insulator surfaces such as Bi$_2$Se$_3$ and Bi$_2$Te$_3$~\cite{Zhang2009Topological,Xia2009Bi2Se3,Chen2009Bi2Te3,Fu2009,hsieh2011nonlinear}, in hexagonal moir\'e platforms with $C_{3v}$ or $D_3$ selection rules~\cite{Wu2018HubbardMoire,Devakul2021MagicTMD,cai2023signatures,park2023observation,zeng2023thermodynamic,xu2023observation}, and in zincblende-type tetrahedral ($T_d$) crystals~\cite{Dresselhaus1955,Winkler2003SpinOrbit}. In these cases the relevant axial representation first appears in the symmetric cubic product of the electric field~\cite{Koster1963,BirPikus1974,Nye1985PhysicalProperties}, so the leading order of orbital-magnetization response is genuinely cubic. 

In this Letter, we develop such a theory from a gauge-invariant finite-momentum cubic Kubo response. The antisymmetric linear-in-$q$ projection that isolates magnetization from transport currents yields three independent dc contributions, a mixed electric-magnetic positional-shift quadrupole $\beta^{(H)}$, a metric or Christoffel-drift term $\beta^{(G)}$, and an orbital-moment octupole transport term $\beta^{(\mathrm{tr})}$. The metric term is the cubic analogue of the quadratic quantum-Christoffel mechanism~\cite{qiang2026quantum,qian2026probing}, and the transport term extends the established semiclassical orbital-moment response. The genuinely new contribution is the mixed quadrupole, which has no quadratic counterpart because it draws on the second-order field-induced orbital moment built from simultaneous electric and magnetic interband mixing. The Kubo construction identifies the mixed-quadrupole and metric contributions only after the Ward-complete finite-$q$ kernel and contact terms are combined. The three contributions transform identically under the point group but remain independent operators with distinct lifetime, frequency, and gate fingerprints, which provide the experimental handles for third-harmonic magneto-optical Kerr spectroscopy. In the continuum benchmark the mixed quadrupole still requires a microscopic magnetic-coupling prescription, so the metric term gives the cleanest closed prediction.

\begin{figure}[t]
\includegraphics[width=0.48\textwidth]{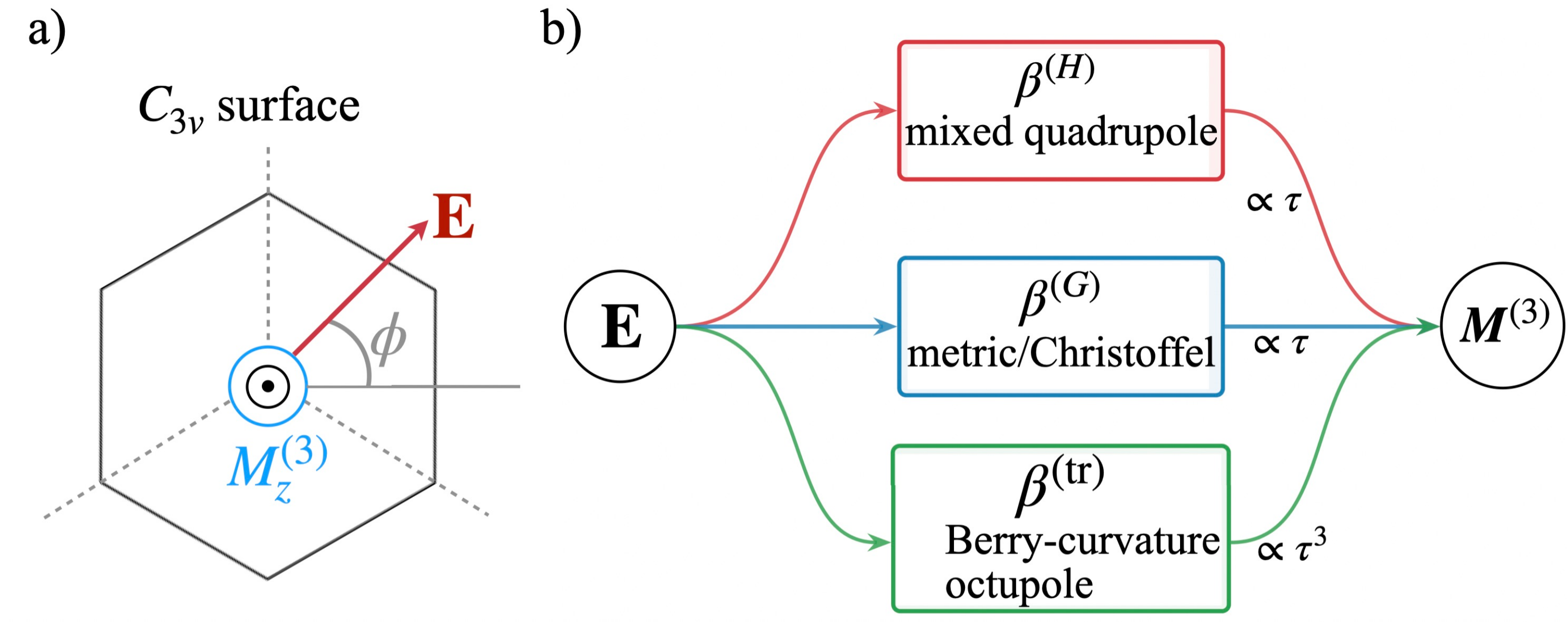}
\caption{\label{figschematic}%
Symmetry filter and three-channel decomposition of cubic orbital magnetization.
(a) In the $C_{3v}$ example, the lower-order axial response is forbidden and the leading signal is the threefold cubic harmonic $\propto E_0^3\cos3\phi$, with the phase fixed by the crystalline axes.
(b) The dc tensor decomposes into three quantum-geometric channels.}
\end{figure}

{\color{blue}\textit{Cubic-order magnetization and the multipole hierarchy}.}\enspace
This selection rule can be seen explicitly on the $C_{3v}$ topological-insulator surface. The out-of-plane magnetization $M_z$ transforms as the axial irrep $A_2$, whereas an in-plane electric field $\mathbf{E}=E_0(\cos\phi,\sin\phi,0)$ transforms as the two-dimensional irrep $E$. Since $\mathrm{Sym}^1(E)=E$ and $\mathrm{Sym}^2(E)=A_1\oplus E$, while $\mathrm{Sym}^3(E)=A_1\oplus A_2\oplus E$, neither the linear nor the quadratic response can generate $M_z$ and the leading allowed term is cubic. This gives the angular fingerprint
$M_z^{(3)}\propto E_0^3\cos3\phi$ or $E_0^3\sin3\phi$, depending on the choice of crystalline axes [Fig.~\ref{figschematic}(a)]~\cite{Koster1963,BirPikus1974,Nye1985PhysicalProperties}. Equivalently, the electric-field-induced orbital magnetization admits the nonlinear expansion
$M_a(\mathbf E)=\chi^{(1)}_{ab}E_b+\chi^{(2)}_{abc}E_bE_c+\chi^{(3)}_{abcd}E_bE_cE_d+\cdots$,
where point-group symmetry selects which response tensors are nonzero~\cite{Koster1963,BirPikus1974,Nye1985PhysicalProperties}. These systems therefore realize a symmetry filter,
$\chi^{(1)}=\chi^{(2)}=0$ but $\chi^{(3)}\neq0$, for the relevant axial magnetization.

When $\chi^{(2)}$ is symmetry allowed, as in orthorhombic $T_d$-phase WTe$_2$ and MoTe$_2$ with $C_{2v}$ point symmetry~\cite{Soluyanov2015,Sun2015Prediction,qiang2026quantum,qian2026probing}, the cubic signal coexists with the quadratic channel and must be isolated by its harmonic, lifetime, and frequency-mixing structure. In the symmetry-filtered $C_{3v}$, $D_3$, and zincblende cases, by contrast, it is the leading magnetization response. The analogy with third-order transport multipoles~\cite{liu2025giant,yu2025quantum,sankar2024experimental,li2024quantum,lai2021third,liu2022berry,zhang2023higher,mandal2024quantum,fang2024quantum,chu2025third,das2023intrinsic,he2024third,lahiri2024nonlinear,xiao2023timereversal} raises the central question of whether orbital magnetization has its own multipole hierarchy and how the individual components can be separated experimentally.

{\color{blue}\textit{Cubic Kubo kernel and $q$-linear projection}}.\enspace
We compute the cubic magnetization from the antisymmetric $q$-linear part of a finite-momentum current response~\cite{xiao2005berry,shi2007orbital,xiao2010berry,SM}, rather than from the strictly uniform $\mathbf q=0$ photocurrent tensor used in nonlinear optical response~\cite{aversa1995nonlinear,sipe2000nonlinear,parker2019diagrammatic,ahn2022riemannian}.
The starting point is the minimally coupled inverse Green function
$G^{-1}(K,A)=i\nu_n+\mu-h(\mathbf{k}-e\mathbf{A}/\hbar)$, where
$K=(\mathbf{k},i\nu_n)$ is the internal momentum-Matsubara label, with
$\mathbf{k}$ the crystal momentum and $\nu_n$ a fermionic Matsubara frequency.
Here $\mu$ is the chemical potential, $h(\mathbf{k})$ represents the Bloch Hamiltonian, and
$\mathbf{A}$ is the electromagnetic vector potential, with the Peierls substitution
taken in its long-wavelength limit. The corresponding generating functional is
\begin{equation}
W[\mathbf{A}]=-T\,{\rm Tr}\ln[-G^{-1}(A)],
\label{eqWfunc}
\end{equation}
with $T$ the temperature and the trace running over $K$ and the internal indices. The current density follows as $j_a(Q)=-(1/V)\,\delta W/\delta A_a(-Q)$, with $V$ the system volume, and three further functional derivatives give the cubic current kernel
\begin{multline}
\widetilde K^R_{a,jkl}(Q,Q_1,Q_2,Q_3) \\
= -\frac{1}{V}\,
\frac{\delta^4 W}{\delta A_a(-Q)\,\delta A_j(Q_1)\,\delta A_k(Q_2)\,\delta A_l(Q_3)}\bigg|_{A=0},
\label{eqkernel}
\end{multline}
where $Q_1,Q_2,Q_3$ are external four-momenta on the input legs, $Q=Q_1+Q_2+Q_3$ enforces energy and momentum conservation, and the ordered retarded branch is obtained after the input ordering and paired frequency labels are fixed~\cite{SM}. Minimal coupling generates, in addition to the paramagnetic vertex
$\Gamma_i^{(1)}=(e/\hbar)\partial_i h$, the diamagnetic vertices
$\Gamma^{(2)}$, $\Gamma^{(3)}$, and $\Gamma^{(4)}$, involving the second through fourth derivatives of $h$. Keeping all box and contact terms gives a 26-term Ward-complete kernel. Omitting the contact terms, or differentiating only the propagator routing while dropping finite-$q$ Peierls vertex derivatives before the Ward-complete $q\to0$ limit, produces spurious longitudinal pieces and a gauge-dependent magnetization projection~\cite{SM}.

The orbital magnetization is obtained from the antisymmetric part of the current that is linear in the external wave vector $\mathbf{q}$. Using
$\mathbf{j}^{(M)}=\nabla\times\mathbf{M}$ together with the frequency-domain relation $E_j(\omega_r)=i\omega_r A_j(\omega_r)$ on each input leg, with $\omega_r$ the frequency on leg $r$, we define
\begin{equation}
\widetilde\beta_{cjkl}(\Omega,\omega_1,\omega_2,\omega_3)
=
\frac{\epsilon_{cab}}{4\omega_1\omega_2\omega_3}
\left[
\frac{\partial\widetilde K^R_{a,jkl}}{\partial q_b}
-
\frac{\partial\widetilde K^R_{b,jkl}}{\partial q_a}
\right]_{\mathbf{q}=0},
\label{eqprojection}
\end{equation}
so that $M_c=\widetilde\beta_{cjkl}E_jE_kE_l$ identifies $\widetilde\beta_{cjkl}$ as the ordered cubic magnetization kernel. Equation~\eqref{eqprojection} is the step that separates a magnetization observable from the usual transport response evaluated at strictly $\mathbf{q}=0$. The calculation is performed at finite external four-momenta, projected with Eq.~\eqref{eqprojection}, and then reduced to the low-frequency single-relaxation-time form. Within the clean noninteracting theory the Ward identities force $\widetilde K^R_{a,jkl}$ to carry an overall factor $\omega_1\omega_2\omega_3$ that cancels the $1/(\omega_1\omega_2\omega_3)$ in Eq.~\eqref{eqprojection}, leaving a finite dc limit for the electric-field tensor~\cite{SM}.

{\color{blue}\textit{Three quantum-geometric contributions}}.\enspace
For the nonmagnetic time-reversal-symmetric case considered here, the dc, single-relaxation-time reduction of Eq.~\eqref{eqprojection} gives a cubic magnetization tensor with three independent quantum-geometric contributions [Fig.~\ref{figschematic}(b)],
\begin{equation}
\widetilde\beta_{ijkl}^{\rm dc}
=
\beta_{ijkl}^{(H)}
+
\beta_{ijkl}^{(G)}
+
\beta_{ijkl}^{({\rm tr})},
\label{eqthree-sector-sum}
\end{equation}
where
\begin{subequations}
\label{eqthree-contributions}
\begin{align}
\beta_{ijkl}^{(H)}
&=
\frac{e^{3}\tau}{\hbar}
\sum_{\nu}\int_{\rm BZ}\frac{d^{d}k}{(2\pi)^{d}}\,
f_{0}(\varepsilon_{\nu})\,
\partial_{l}H_{ijk}^{\nu},
\label{eqbetaH-main}
\\
\beta_{ijkl}^{(G)}
&=
\frac{e^{3}\tau}{2\hbar}
\sum_{\nu}\int_{\rm BZ}\frac{d^{d}k}{(2\pi)^{d}}\,
(\partial_{l}m_{i}^{\nu})\,
\mathcal{G}_{jk}^{\nu}\,
f_{0}'(\varepsilon_{\nu}),
\label{eqbetaG-main}
\\
\beta_{ijkl}^{({\rm tr})}
&=
\frac{e^{3}\tau^{3}}{\hbar^{3}}
\sum_{\nu}\int_{\rm BZ}\frac{d^{d}k}{(2\pi)^{d}}\,
f_{0}(\varepsilon_{\nu})\,
\partial_{j}\partial_{k}\partial_{l}m_{i}^{\nu}.
\label{eqbetatr-main}
\end{align}
\end{subequations}
Here $\tau$ is the relaxation time, $\varepsilon_\nu$ is the band energy for the band $\nu$ and $f_{0}(\varepsilon_\nu)$ is the equilibrium Fermi function with $f_{0}'\equiv\partial_{\varepsilon}f_{0}$.  The intrinsic orbital magnetic moment $m_{i}^{\nu}$ is
\begin{equation}
m_{i}^{\nu}
=
\frac{e}{2\hbar}\epsilon_{iab}\,
{\rm Im}\,
\langle\partial_{a}u_{\nu}|
(\varepsilon_{\nu}-h)
|\partial_{b}u_{\nu}\rangle,
\label{eqm-def}
\end{equation}
the gap-weighted quantum metric $\mathcal{G}_{jk}^{\nu}$ is
\begin{equation}
\mathcal{G}_{jk}^{\nu}
=
2\sum_{\mu\neq\nu}
\frac{{\rm Re}\,A_{\nu\mu}^{j}A_{\mu\nu}^{k}}
{\varepsilon_{\nu}-\varepsilon_{\mu}},
\label{eqGmetric-def}
\end{equation}
and the symmetrized mixed electric-magnetic positional-shift contribution to the
second-order field-induced orbital moment is
\begin{equation}
\delta m_{\nu,i}^{(2)}
=
e^{2}H_{ijk}^{\nu}E_{j}E_{k},
\qquad
H_{ijk}^{\nu}
=
-\frac{1}{2}
\left(
\mathcal Q_{jki}^{\nu}
+
\mathcal Q_{kji}^{\nu}
\right),
\label{eqH-def}
\end{equation}
where $A_{\nu\mu}^{a}=i\langle u_{\nu}|\partial_{a}u_{\mu}\rangle$ is the interband
Berry connection and $u_\nu$ the cell-periodic Bloch eigenstate. The explicit
gauge-covariant construction of $\mathcal Q_{\alpha\beta\gamma}^{\nu}$ from
electric and magnetic interband mixing amplitudes is given in the Supplemental
Material~\cite{SM}.

Equations~\eqref{eqthree-contributions} have a simple multipole interpretation. The $\beta^{(H)}$ contribution is the occupied-state dipole of a mixed electric-magnetic quadrupolar correction to the local orbital moment. The $\beta^{(G)}$ term is a Fermi-surface metric drift, or equivalently
$\beta^{(G)}_{ijkl}=-(e^{3}\tau/2\hbar)\sum_{\nu}\int_{\rm BZ} m_i^\nu
\partial_l[\mathcal G_{jk}^{\nu}f_0'(\varepsilon_\nu)]$,
so the derivative $\partial_l\mathcal G_{jk}^{\nu}$ can be written as
$\Gamma^{(\mathcal G),\nu}_{j,lk}+\Gamma^{(\mathcal G),\nu}_{k,lj}$, with
$\Gamma^{(\mathcal G),\nu}_{c,ab}
=\frac12(\partial_a\mathcal G_{bc}^{\nu}
+\partial_b\mathcal G_{ac}^{\nu}
-\partial_c\mathcal G_{ab}^{\nu})$. In a two-band model the band-normalized metric is $\mathcal G^\nu_{jk}=2g^\nu_{jk}/\Delta_{\nu\bar\nu}$, with $g^\nu_{jk}$ the quantum metric and $\Delta_{\nu\bar\nu}$ the interband gap, so this reorganization extracts the gap-weighted Christoffel structure but leaves an additional gap-gradient term $-2(\partial_l\Delta_{\nu\bar\nu})g^\nu_{jk}/\Delta_{\nu\bar\nu}^2$ that vanishes only when the interband gap is momentum independent~\cite{SM}. The transport contribution is the orbital-moment octupole. In a two-band model, using
$m_i^\nu=(e\Delta_{\nu\bar\nu}/2\hbar)\Omega_i^\nu$ with $\Omega_i^\nu$ the Berry curvature, it becomes a gap-weighted Berry-curvature octupole, reducing to a pure Berry-curvature octupole only when the gap is momentum independent.

{\color{blue}\textit{Contribution diagnostics}.}\enspace
Although point-group symmetry cannot separate the three contributions, their dynamical fingerprints can. In the low-frequency single-relaxation-time reduction of Eq.~\eqref{eqprojection}, the mixed-quadrupole and metric contributions are both linear in $\tau$, whereas the orbital-moment-octupole transport contribution scales as $\tau^3$~\cite{SM}. Their finite-frequency dispersion also differs. For an ordered input sequence $(j,\omega_1),(k,\omega_2),(l,\omega_3)$, the mixed-quadrupole contribution relaxes at the last input frequency $\omega_3$, the metric contribution at the output frequency $\Omega=\omega_1+\omega_2+\omega_3$, and the transport contribution at all three of $\omega_3$, $\omega_2+\omega_3$, and $\Omega$. The physical response is obtained by symmetrizing over the paired input labels~\cite{SM}. For a degenerate THG drive, this corresponds to crossovers at $\omega\tau\sim1$ for the mixed quadrupole, $3\omega\tau\sim1$ for the metric term, and at $\omega\tau$, $2\omega\tau$, and $3\omega\tau\sim1$ for the transport contribution.

In the dc or quasi-static regime, the Drude factors $(1-i\omega_r\tau)^{-1}$ on each leg approach unity, and the scalar $C_{3v}$ coefficient defined in Eq.~\eqref{eqC3v-kappa} reduces to $\bar\chi = A_H\tau + A_G\tau + A_{\rm tr}\tau^3$, with $A_H$, $A_G$, and $A_{\rm tr}$ the lifetime-independent band-geometry amplitudes~\cite{SM}.
A log-log plot of $|\bar\chi|$ versus $\tau$ therefore crosses from slope $1$ in the disorder-dominated geometric regime to slope $3$ in the clean transport-dominated regime, unless one of the microscopic coefficients is accidentally suppressed~\cite{SM}.
This lifetime scaling separates the two geometric contributions collectively from the transport contribution, while gate and frequency diagnostics distinguish the mixed quadrupole from the metric term. This separation refers to the clean Ward-complete kernel and can be complicated by extrinsic scattering, since side-jump terms are typically $\tau$-independent and could mimic geometric scaling in the moderate-disorder window, while skew-scattering adds process-dependent powers of $\tau$~\cite{lai2021third,liu2022berry,he2024third,du2021nonlinearreview}. The diagnostic is therefore strongest when combined with the gate-tunable cutoff-independent fingerprint of the metric contribution and with two-color frequency scans.
In particular, for the weak-warping two-band $C_{3v}$ model analyzed below, the metric contribution scales as $\mu^{-2}$ with the gate-controlled chemical potential, which provides a direct electrostatic test~\cite{SM}.
For $k_BT\ll|\mu|$ and away from band singularities, the remaining temperature dependence enters mainly through Fermi-surface smearing and $\tau(T)$. With an independent calibration of $\tau(T)$, temperature or disorder scans can factor out the lifetime powers~\cite{Tian2009,Du2019}.

{\color{blue}\textit{$C_{3v}$ benchmark and lattice completion}}.\enspace
To illustrate the three diagnostics, we use the standard hexagonally warped topological-insulator surface Hamiltonian~\cite{Fu2009,SM},
\begin{align}
H_{C_{3v}}(\mathbf{k})
&=
\left(\frac{k_x^2+k_y^2}{2m^\ast}-\mu\right)\sigma_0
+v(k_x\sigma_y-k_y\sigma_x)
\nonumber\\
&\quad
+\lambda(k_x^3-3k_xk_y^2)\sigma_z ,
\label{eqC3v-benchmark}
\end{align}
where $m^\ast$ is the band mass, $v$ is the Dirac velocity, $\lambda$ is the hexagonal-warping strength, $\sigma_0$ is the identity and $\sigma_{x,y,z}$ are the Pauli matrices, the Fermi level is set to zero, and $\mu$ is the gate-controlled chemical-potential offset. The cubic warping
$w(\mathbf{k})=k^3\cos3\varphi_{\mathbf{k}}$, with $\varphi_{\mathbf{k}}$ the polar angle of $\mathbf{k}$,
is the lowest-order $C_{3v}$ anisotropy compatible with time reversal in this surface model~\cite{Fu2009}, and it breaks inversion and supplies the anisotropy needed for the out-of-plane orbital magnetization. For the crystalline orientation in Eq.~\eqref{eqC3v-benchmark}, the $C_{3v}$ axial rank-four tensor reduces, for in-plane fields, to a single scalar coefficient~\cite{Koster1963,BirPikus1974,Nye1985PhysicalProperties,SM},
\begin{equation}
M_z^{(3)}
=
\bar\chi E_0^3\cos3\phi ,
\label{eqC3v-kappa}
\end{equation}
where $\mathbf{E}=E_0(\cos\phi,\sin\phi,0)$ and
$\bar\chi=\bar\chi_H+\bar\chi_G+\bar\chi_{\rm tr}$. 

In the weak-warping regime $|\lambda|k_F^2\ll v$, with $k_F$ the Fermi wavevector, and at zero temperature, the Fermi-surface integral of Eq.~\eqref{eqbetaG-main} gives
\begin{equation}
\bar\chi_G^{T=0}
=
\frac{e^4\tau\lambda}{32\pi\hbar^2\mu^2}
+O(\lambda^3),
\label{eqchibarG-closed}
\end{equation}
which provides the gate-tunable $\mu^{-2}$ fingerprint of the metric/Christoffel channel. The continuum model does not determine all remaining coefficients in a cutoff-independent way. The transport coefficient $\bar\chi_{\rm tr}$ depends on the ultraviolet regularization of the continuum integral, while the mixed-quadrupole coefficient $\bar\chi_H$ requires the interband magnetic matrix element, or equivalently a microscopic magnetic-coupling prescription. In this sense, the metric channel is the cleanest theoretical prediction of the framework, in that the other two contributions are well-defined gauge-invariant operators but their amplitudes are model-dependent. A triangular-lattice completion of Eq.~\eqref{eqC3v-benchmark}, described in the Supplemental Material, fixes these ultraviolet and magnetic-coupling ambiguities while reproducing the same $C_{3v}$ Hamiltonian near $\Gamma$~\cite{SM}.

\begin{figure}[t]
\includegraphics[width=0.5\textwidth]{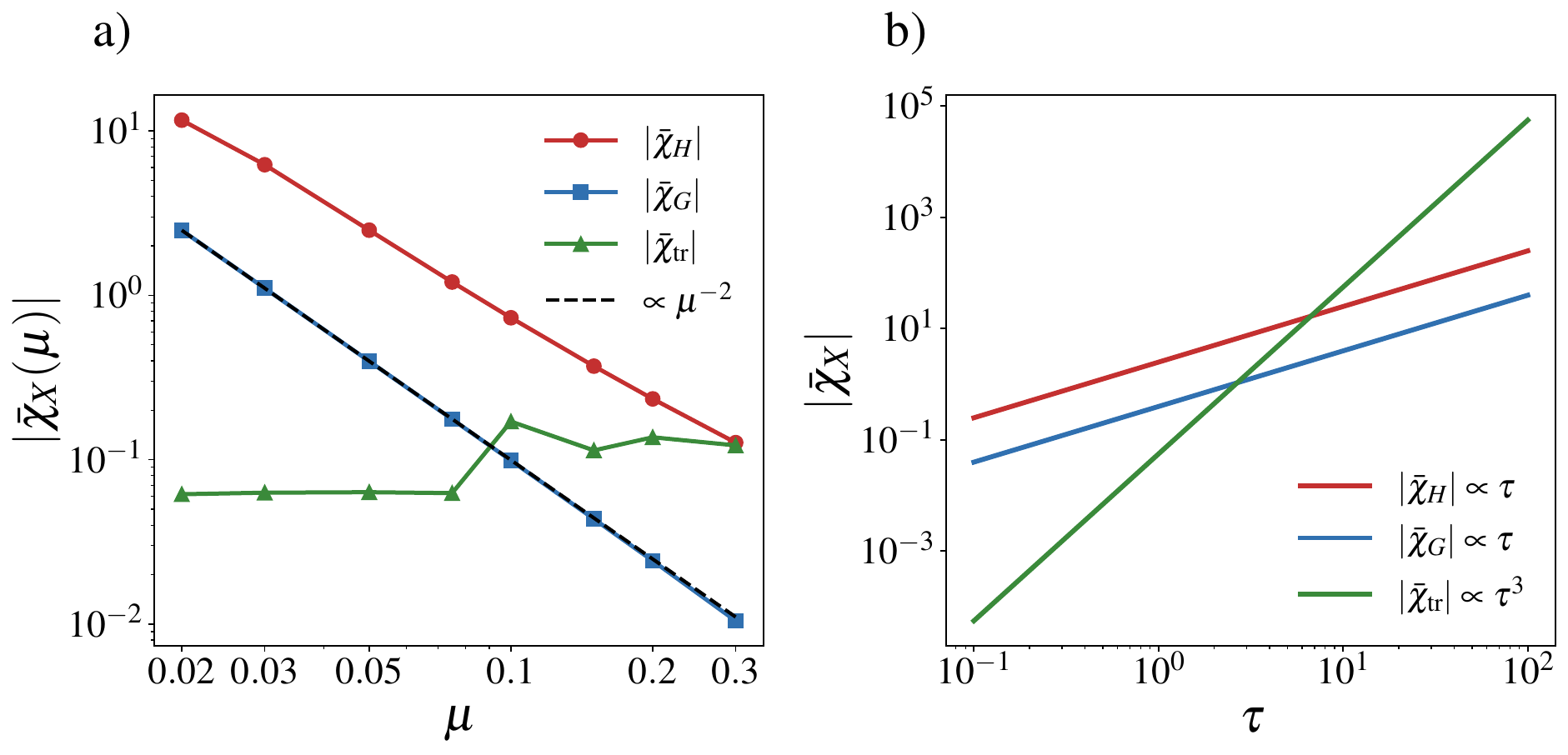}
\caption{\label{figbenchmark}%
Contribution-resolved cubic magnetization in the $C_{3v}$ model of Eq.~\eqref{eqC3v-benchmark}. (a)~The three coefficients $|\bar\chi_{H}|$, $|\bar\chi_{G}|$, and $|\bar\chi_{\mathrm{tr}}|$ as functions of chemical potential $\mu$, with the dashed line the $\mu^{-2}$ prediction of Eq.~\eqref{eqchibarG-closed} for $|\bar\chi_{G}|$. The ordering of the three magnitudes shown here is not universal and depends on microscopic details of the material. (b)~The same coefficients as functions of relaxation time $\tau$, showing the $\tau^1$ scaling of the geometric contributions and the $\tau^3$ scaling of the transport contribution from Eqs.~\eqref{eqthree-contributions}.}
\end{figure}

Figure~\ref{figbenchmark} confirms the two predicted fingerprints, the gate-tunable $\mu^{-2}$ law of Eq.~\eqref{eqchibarG-closed} for the metric term and the $\tau^1$ versus $\tau^3$ separation of the geometric and transport contributions. Their relative magnitudes are not universal and depend on microscopic details of the material.

{\color{blue}\textit{THG-MOKE detection and material platforms}}.\enspace
The finite-frequency cubic magnetization can be accessed by third-harmonic magneto-optical Kerr spectroscopy. For a degenerate in-plane drive at frequency $\omega$, the emitted magnetization component at $\Omega=3\omega$ is
\[
M_z^{(3)}(3\omega)
=
\left[
\bar\chi_H+\bar\chi_G+\bar\chi_{\rm tr}
\right]_{(3\omega,\omega,\omega,\omega)}
E_0^3\cos3\phi
\]
in the crystalline convention of Eq.~\eqref{eqC3v-benchmark}. The corresponding Kerr rotation and ellipticity are proportional to the magneto-optical source at $3\omega$,
\[
\theta_K(3\omega)+i\eta_K(3\omega)
=
F(3\omega,n_s,d,\sigma_{\rm opt},\ldots)\,
M_z^{(3)}(3\omega),
\]
where the optical factor $F$ is fixed by the transfer matrix of the film, substrate, and optical boundary conditions. In the idealized normal-incidence bulk limit this reduces schematically to the familiar estimate
$\theta_K+i\eta_K\simeq (4\pi/c)M_z^{(3)}/(n^2-1)$, with $c$ the speed of light and $n$ the refractive index, but quantitative extraction in thin films or surface-state systems should use the full optical geometry~\cite{argyres1955theory,bennett1965faraday,zvezdin1997modern,SM}. This is the cubic analogue of the SHG-MOKE strategy of Ref.~\cite{qian2026probing}. Using Eq.~\eqref{eqchibarG-closed} with $\lambda\sim250$~eV\,\AA$^3$, $\mu\sim0.1$~eV, and $\tau\sim10^{-13}$~s gives a sheet response $|\bar\chi_G|\simeq9.2\times10^{-30}\,{\rm A}({\rm V/m})^{-3}$. For $E_0\sim10^4$~V\,cm$^{-1}$ this gives $M_z^{(3)}\simeq9.2\times10^{-12}$~A, equivalent to about $1.0\times10^{-6}\,\mu_B$\,nm$^{-2}$ for the metric channel at $3\omega$. The corresponding Kerr scale is therefore a demanding small-signal target and should be estimated with the full optical transfer matrix. Since the signal scales as $E_0^3$, a stronger pulsed THz drive with $E_0\sim10^5$~V\,cm$^{-1}$ would enhance the same estimate by $10^3$ and give $M_z^{(3)}\sim10^{-3}\,\mu_B$\,nm$^{-2}$.

THG-MOKE measures a coherent sum of all active cubic contributions rather than a contribution-resolved observable, and even the frequency rolloffs become superposed under a single-color drive. For a degenerate pump the symmetrization over the three input legs averages the ordered $\omega_3$, $\omega_2+\omega_3$, and $\Omega$ Drude denominators, so the per-contribution rolloffs derived in the Supplemental Material are recovered only in a non-degenerate two-color experiment. A three-way decomposition can nevertheless be obtained by combining the diagnostics derived above. Disorder or temperature scans calibrated by an independent $\tau$ measurement separate geometric from transport scaling, electrostatic gating tests Eq.~\eqref{eqchibarG-closed}, and non-degenerate two-color measurements resolve the distinct ordered-kernel rolloffs~\cite{SM}.

Possible backgrounds, including ordinary optical third-harmonic generation, inverse-Faraday effects, drive-induced Oersted fields, and bolometric heating, should be controlled rather than assumed absent. The usual inverse-Faraday source is quadratic in the optical field and vanishes for a linearly polarized drive~\cite{pershan1966theoretical,kimel2005ultrafast}, whereas the cubic magnetization studied here is finite for linear polarization and carries the $\cos3\phi$ or $\sin3\phi$ angular fingerprint of $C_{3v}$. Comparing linear and circular polarizations, together with the $C_{3v}$ harmonic, phase-sensitive Kerr detection, the cubic power law in the drive field, gate tracking, and two-color frequency dependence, gives discriminants that a substrate-dominated, thermal, or purely optical background is not expected to reproduce simultaneously~\cite{SM}.

The most direct platforms are $C_{3v}$ topological-insulator surfaces such as Bi$_2$Se$_3$ and Bi$_2$Te$_3$, where hexagonal warping realizes Eq.~\eqref{eqC3v-benchmark} and surface gating can test Eq.~\eqref{eqchibarG-closed}~\cite{Fu2009,hsieh2011nonlinear}. Hexagonal moir\'e heterobilayers provide a tunable setting in which twist angle, displacement field, and carrier density can modify both the warping and the active Fermi surfaces, while flat bands may enhance the orbital moment and quantum metric near moir\'e Dirac points~\cite{Wu2018HubbardMoire,Devakul2021MagicTMD,cai2023signatures,park2023observation,zeng2023thermodynamic,xu2023observation,torma2022superconductivity,tian2023evidence,tanaka2025superfluid}. Zincblende-type tetrahedral crystals supply a three-dimensional realization of the same cubic-leading axial response~\cite{Dresselhaus1955,Winkler2003SpinOrbit}. Orthorhombic $T_d$-phase WTe$_2$ and MoTe$_2$ are useful comparison platforms because existing nonlinear Kerr and SHG-MOKE methods can be extended to the $3\omega$ channel, even though quadratic magnetization channels are symmetry allowed~\cite{qiang2026quantum,qian2026probing,wu2023extrinsic,ye2024nonlinear}.

{\color{blue}\textit{Discussion}.}\enspace
The finite-$q$ Kubo construction developed here is not tied to the $C_{3v}$ model. For any noncentrosymmetric point group, the antisymmetric $q$-linear projection separates the magnetization part of the cubic current response from the uniform transport response. Point-group symmetry determines which components of the resulting axial rank-four tensor are allowed, but it does not distinguish the microscopic origins contained in Eq.~\eqref{eqthree-contributions}. Those must be identified dynamically.

Magnetic noncentrosymmetric crystals enlarge this structure. Once time reversal is broken, an additional low-frequency contribution
$\beta^{(\mathcal M\delta f)}\equiv\delta m^{(1)}\delta f^{(2)}$
can survive, with $\delta m^{(1)}$ the linear field-induced correction to the band orbital moment and $\delta f^{(2)}$ the second-order correction to the distribution function. The contributing band objects are odd under $\mathbf{k}\to-\mathbf{k}$ when time reversal is unbroken, so $\beta^{(\mathcal M\delta f)}$ vanishes in nonmagnetic crystals but can be finite in noncentrosymmetric magnets. The resulting symmetry classification of centrosymmetric, time-reversal-symmetric noncentrosymmetric, $PT$-symmetric noncentrosymmetric magnetic, and generic noncentrosymmetric magnetic classes is summarized in Table~S2 of the Supplemental Material~\cite{SM}. A particularly useful limiting case is provided by $PT$-symmetric noncentrosymmetric magnets. For nondegenerate isolated bands, $PT$ symmetry enforces $m_i^\nu=0$ and removes the mixed-quadrupole moment correction, so the mixed-quadrupole, metric, and transport contributions vanish while $\beta^{(\mathcal M\delta f)}$ may remain. Thus this class can isolate the magnetic contribution, although degenerate $PT$ bands require the corresponding non-Abelian trace formulation.

The framework admits several extensions. The scaling analysis used a scalar relaxation time inserted after the Ward-complete clean kernel was organized, and quantitative material predictions should add energy-dependent lifetimes, vertex corrections, side-jump, and skew-scattering mechanisms known from nonlinear Hall transport~\cite{nandy2019symmetry,xiao2019theory,du2021quantum}. These renormalize the crossover and add extrinsic pieces, but the clean geometric decomposition provides the reference structure. The theory connects naturally to orbitronics, where recent experiments access current-induced orbital magnetization through orbital Hall and orbital Edelstein effects~\cite{choi2023observation,lyalin2023magnetooptical,hayashi2024observation,wang2024orbital,ye2024nonlinear,go2021orbitronics}, the present response being their cubic quantum-geometric counterpart. In correlated metals, the band objects $H_{ijk}^{\nu}$, $\mathcal G_{jk}^{\nu}$, and $m_i^\nu$ should be replaced by interacting Green-function or quasiparticle analogues, paralleling the interacting theory of equilibrium orbital magnetization~\cite{thonhauser2005orbital,ceresoli2006orbital}. Cubic orbital magnetization therefore probes Bloch-state geometry beyond the Berry-curvature-dipole level.

{\color{blue}\textit{Acknowledgments}.}\enspace This work was supported by the Research Council
of Norway through its Centers of Excellence funding
scheme, Project No. 353919  and  Project No. 361800 “QTransMag.”

\bibliography{Refs}

\clearpage
\onecolumngrid
\setcounter{secnumdepth}{3}
\setcounter{section}{0}
\setcounter{equation}{0}
\setcounter{figure}{0}
\setcounter{table}{0}
\renewcommand{\thesection}{S\arabic{section}}
\renewcommand{\theequation}{S\arabic{equation}}
\renewcommand{\thefigure}{S\arabic{figure}}
\renewcommand{\thetable}{S\arabic{table}}

\begin{center}
{\large\bfseries Supplemental Material:\\[2pt]
Three Quantum-Geometric Contributions to Cubic Orbital Magnetization}\\[8pt]
Tohid Farajollahpour\\[2pt]
{\small\itshape Department of Physics, Norwegian University of Science and Technology (NTNU), NO-7491 Trondheim, Norway}\\
{\small\itshape Department of Physics, Brock University, St.\ Catharines, Ontario L2S 3A1, Canada}
\end{center}
\vspace{6pt}

In this Supplemental Material, we first construct the clean, noninteracting third-order current kernel from minimal coupling and show how its antisymmetric $q$-linear part yields the cubic magnetization. We then derive the gauge-covariant static moment tensor, develop the ordered low-frequency expansion, reduce the diagrammatic kernel to the three dc contributions, summarize the symmetry constraints, and give continuum and lattice models together with experimental diagnostics.

\section{Minimal coupling, vertices, and the full third-order kernel}
\label{seckubo-kernel}
The cubic current kernel is built from two standard ingredients: the
minimal-coupling prescription for Bloch electrons in an electromagnetic
field, and the generating-functional formulation of linear and nonlinear
response~\cite{kubo1957statistical,mahan2000many,parker2019diagrammatic}.
Let $h(\bk)$ denote the single-particle Bloch Hamiltonian, a matrix in the
internal orbital, band, and spin indices, with $\bk$ the crystal momentum.
The electromagnetic field enters through a vector potential
$\bA(\mathbf r,\tau)$, treated as a classical external source, where
$\mathbf r$ is position and $\tau$ is imaginary (Matsubara) time. In Fourier
space its Cartesian components are $A_a(Q)$, with $a\in\{x,y,z\}$ a spatial
direction and $Q\equiv(\bq,\ii\Omega_m)$ a bosonic four-momentum that
combines the transferred wavevector $\bq$ with a bosonic Matsubara frequency
$\Omega_m$. Minimal coupling is the Peierls shift
$\bk\to\bk-(e/\hbar)\bA$, in which $e>0$ is the elementary charge and
$\hbar$ the reduced Planck constant, so that the field-dressed Bloch
Hamiltonian reads
\begin{equation}
h_A(\bk)=h\!\left(\bk-\frac{e}{\hbar}\bA\right).
\label{eqminimal-coupling-supp}
\end{equation}
In Eq.~\eqref{eqminimal-coupling-supp} the bare symbol $\bA$ is a uniform-field shorthand for the general source $A_a(Q)$, whose four-momentum dependence is reinstated when the vertices are assigned incoming momenta below.  The single-particle inverse Green function is
\begin{equation}
G^{-1}(K,A)=\ii\nu_n+\mu-h_A(\bk),
\qquad K\equiv(\bk,\ii\nu_n),
\label{eqGinv-supp}
\end{equation}
where $\mu$ is the chemical potential and $\nu_n$ is a fermionic Matsubara frequency. The grand-potential functional is
\begin{equation}
W[A]=-T\,\Tr\ln[-G^{-1}(A)],
\label{eqW-supp}
\end{equation}
where $T$ denotes temperature in energy units. Equation~\eqref{eqW-supp} generates the current-response functions. With the convention
\begin{align}
j_a(Q)\equiv -\frac{1}{V}\frac{\delta W[A]}{\delta A_a(-Q)},
\end{align}
variation of Eq.~\eqref{eqW-supp} gives
\begin{align}
\delta W[A]=-T\,\Tr\!\left[G[A]\,\delta G^{-1}[A]\right].
\end{align}
Therefore the current density can be written as
\begin{equation}
j_a(Q)=\frac{T}{V}\Tr\left[G[A]X_a(Q,A)\right],
\qquad X_a(Q,A)=\frac{\delta G^{-1}[A]}{\delta A_a(-Q)}.
\label{eqjGX-supp}
\end{equation}
where $V$ is the crystal volume and $\Tr$ runs over momentum, fermionic Matsubara frequency, and internal orbital, band, and spin indices. The source vertex $X_a(Q,A)$ is the functional derivative of $G^{-1}$ with respect to $A_a(-Q)$. At $A=0$ it reduces to the usual paramagnetic current vertex $\Gamma^{(1)}_a=(e/\hbar)\partial_a h$. Higher derivatives generate the diamagnetic/contact vertices required by nonlinear Ward identities~\cite{aversa1995nonlinear,parker2019diagrammatic}. The higher functional derivatives of $G^{-1}$ with respect to the source
field define the contact (diamagnetic) vertices are given by 
\begin{equation}
X_{ab}=\frac{\delta^2G^{-1}}{\delta A_a\delta A_b},
\qquad
X_{abc}=\frac{\delta^3G^{-1}}{\delta A_a\delta A_b\delta A_c},
\qquad
X_{abcd}=\frac{\delta^4G^{-1}}{\delta A_a\delta A_b\delta A_c\delta A_d}.
\label{eqX-higher-supp}
\end{equation}
Throughout this section, $a,j,k,l\in\{x,y,z\}$ denote Cartesian directions and $\partial_i\equiv\partial/\partial k_i$. At $A=0$ these become the current vertices
\begin{align}
\Gamma^{(1)}_i&=X_i\big|_{A=0}=\frac{e}{\hbar}\partial_i h,
\label{eqGamma1-supp}\\
\Gamma^{(2)}_{ij}&=X_{ij}\big|_{A=0}=-\frac{e^2}{\hbar^2}\partial_i\partial_j h,
\label{eqGamma2-supp}\\
\Gamma^{(3)}_{ijk}&=X_{ijk}\big|_{A=0}=\frac{e^3}{\hbar^3}\partial_i\partial_j\partial_k h,
\label{eqGamma3-supp}\\
\Gamma^{(4)}_{ijkl}&=X_{ijkl}\big|_{A=0}=-\frac{e^4}{\hbar^4}\partial_i\partial_j\partial_k\partial_l h.
\label{eqGamma4-supp}
\end{align}
Using $
\delta G=-G(\delta G^{-1})G$, 
one finds the first and second functional derivatives. For compactness we write $\delta_j\equiv\delta/\delta A_j(-Q_1)$, $\delta_k\equiv\delta/\delta A_k(-Q_2)$, and $\delta_l\equiv\delta/\delta A_l(-Q_3)$
\begin{align}
\delta_j j_a&=\frac{T}{V}\Tr\left[-GX_jGX_a+GX_{ja}\right],
\label{eqfirst-derivative-supp}\\
\delta_k\delta_j j_a&=\frac{T}{V}\Tr\Big[GX_kGX_jGX_a+GX_jGX_kGX_a-GX_{jk}GX_a
\nonumber\\
&\qquad\qquad\qquad\qquad -GX_jGX_{ka}-GX_kGX_{ja}+GX_{jka}\Big].
\label{eqsecond-derivative-supp}
\end{align}
Differentiating once more with respect to $A_l$ yields the exact ordered cubic kernel in compact operator form,
\begin{equation}
\delta_l\delta_k\delta_j j_a=\frac{T}{V}\Tr\left[\calB_{a,jkl}+\calC_{a,jkl}+\calD_{a,jkl}+\calQ_{a,jkl}\right],
\label{eqthird-derivative-compact-supp}
\end{equation}

The cubic kernel in Eq.~\eqref{eqthird-derivative-compact-supp} is an
ordered response, with the three incoming fields carrying the ordered
label-frequency pairs $(j,Q_1)$, $(k,Q_2)$, and $(l,Q_3)$. The same ordering
convention applies to the cubic magnetization kernel
$\widetilde\beta_{ijkl}(\Omega,\omega_1,\omega_2,\omega_3)$ introduced below,
where $\Omega=\omega_1+\omega_2+\omega_3$ is the output frequency. The
physical tensor that enters the symmetry analysis follows from symmetrizing
over simultaneous permutations of these three field-index/frequency pairs.
Writing $(j_1,j_2,j_3)\equiv(j,k,l)$,
\begin{equation}
\beta^{\mathrm{phys}}_{ijkl}(\Omega,\omega_1,\omega_2,\omega_3)
=
\frac{1}{3!}\sum_{\pi\in S_3}
\widetilde\beta_{i j_{\pi_1}j_{\pi_2}j_{\pi_3}}
\bigl(\Omega,\omega_{\pi_1},\omega_{\pi_2},\omega_{\pi_3}\bigr).
\label{eqbeta-phys-sym-supp}
\end{equation}
In the static or fully degenerate limit this reduces to ordinary
symmetrization over the last three Cartesian indices, the convention adopted
in the point-group component tables. For nondegenerate frequency mixing the
intrinsic permutation symmetry instead acts on the paired labels
$(j,\omega_1)$, $(k,\omega_2)$, and $(l,\omega_3)$. Equivalently, one may
retain the ordered kernels $\widetilde\beta$ and sum those ordered
contributions that share the same output frequency $\Omega$;
Eq.~\eqref{eqbeta-phys-sym-supp} is the convention in which this sum is
absorbed into $\beta^{\mathrm{phys}}$.

The four vertex classes entering Eq.~\eqref{eqthird-derivative-compact-supp} are
\begin{align}
\calB_{a,jkl}
&=-\sum_{\pi\in S_3}GX_{\pi_3}GX_{\pi_2}GX_{\pi_1}GX_a,
\label{eqBclass-supp}\\
\calC_{a,jkl}
&=\big(GX_{jk}GX_lGX_a+GX_lGX_{jk}GX_a+GX_{jl}GX_kGX_a\big)
\nonumber\\
&\quad+\big(GX_kGX_{jl}GX_a+GX_{kl}GX_jGX_a+GX_jGX_{kl}GX_a\big)
\nonumber\\
&\quad+\big(GX_jGX_kGX_{al}+GX_kGX_jGX_{al}+GX_jGX_lGX_{ak}\big)
\nonumber\\
&\quad+\big(GX_lGX_jGX_{ak}+GX_kGX_lGX_{aj}+GX_lGX_kGX_{aj}\big),
\label{eqCclass-supp}\\
\calD_{a,jkl}
&=-\big(GX_{jk}GX_{al}+GX_{jl}GX_{ak}+GX_{kl}GX_{aj}\big)
\nonumber\\
&\quad-\big(GX_{jkl}GX_a+GX_jGX_{akl}+GX_kGX_{ajl}+GX_lGX_{ajk}\big),
\label{eqDclass-supp}\\
\calQ_{a,jkl}&=GX_{ajkl}.
\label{eqQclass-supp}
\end{align}
Here $S_3$ permutes the three incoming labels $(j,k,l)$ in the box class. The Ward-complete cubic kernel contains $
6+12+3+4+1=26$ terms. They consist of 6 box terms with four rank-one vertices, 12 mixed-contact terms with one $\Gamma^{(2)}$, 3 double-contact terms, 4 cubic-contact terms, and 1 quartic-contact term. The detailed list in Eqs.~\eqref{eqBclass-supp} through \eqref{eqQclass-supp} is kept because these contact vertices are not optional. They are generated by differentiating the minimally coupled $G^{-1}(A)$ and are required by the nonlinear Ward identities. This is the nonlinear analogue of the gauge-invariant organization familiar from optical-response theory~\cite{aversa1995nonlinear,parker2019diagrammatic}. A useful algebraic check is obtained by differentiating the gauge invariance of the source functional, $W[A+\nabla\chi]=W[A]$. In four-vector notation the clean full kernel obeys
\begin{equation}
Q_\mu\,\widetilde K^{\mu,jkl}(Q,Q_1,Q_2,Q_3)=0,
\label{eqward-output-supp}
\end{equation}
with analogous Ward identities on the three incoming legs. In the static long-wavelength limit this reduces to the requirement that a spatially uniform vector potential, which is a pure gauge on a simply connected sample, cannot produce a physical current. The same cancellation can be traced term by term from the vertex identities generated by the Peierls substitution, e.g.,
\begin{equation}
q_a\Gamma^{(1)}_a(\bk,\bq)=\frac{e}{\hbar}\big[h(\bk+\bq)-h(\bk)\big]+\order{q^2},
\label{eqvertex-ward-supp}
\end{equation}
whose next derivatives produce the longitudinal parts of $\Gamma^{(2)}$, $\Gamma^{(3)}$, and $\Gamma^{(4)}$. The box class alone leaves uncanceled longitudinal pieces, the mixed-, double-, cubic-, and quartic-contact families cancel them so that Eq.~\eqref{eqward-output-supp} holds before any low-frequency or relaxation-time reduction is made.

\begin{figure}[t]
\centering
\includegraphics[width=0.58\textwidth]{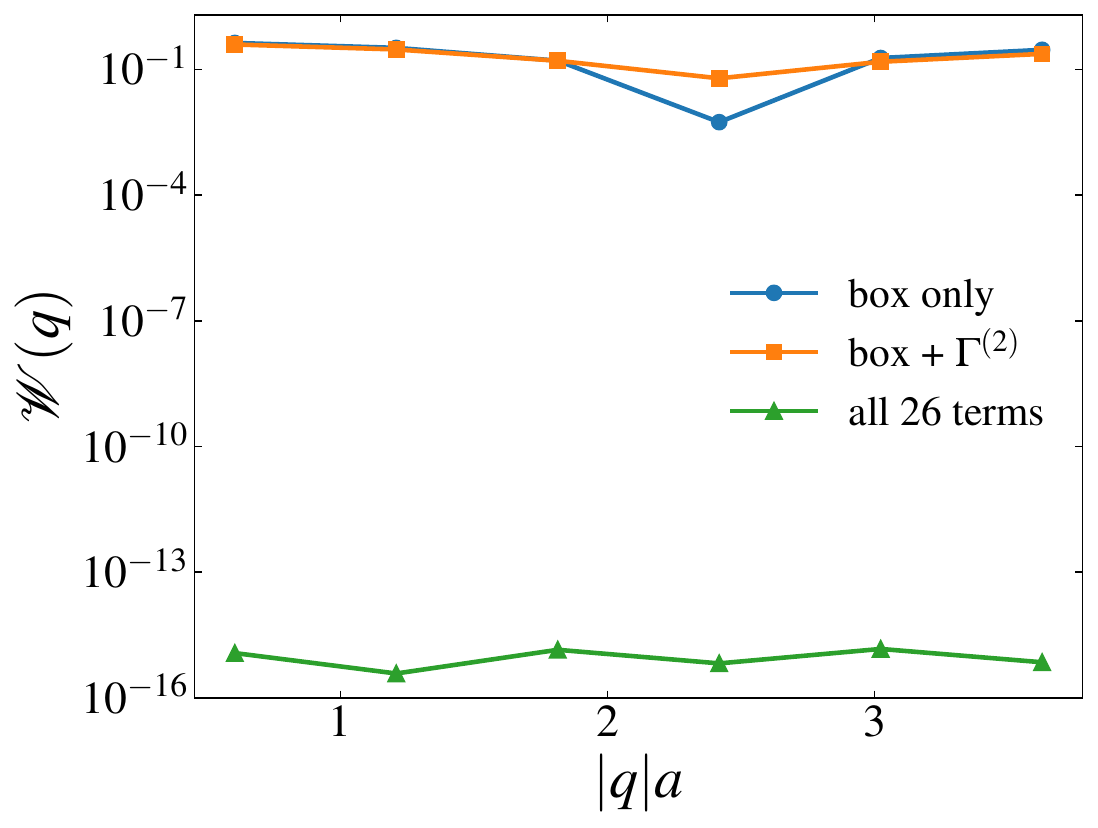}
\caption{
Output-leg Ward-identity check for the cubic current kernel. We plot the normalized longitudinal residual
$
\mathcal W(q)=
|q_a\widetilde K_{a,jkl}(q)|/(|q|K_{\rm box}^{\rm max})
$
for the triangular-lattice Peierls regularization of the $C_{3v}$ model, with $K_{\rm box}^{\rm max}=\max_a|\widetilde K^{\rm box}_{a,jkl}|$. Parameters are $a=e=\hbar=v=m^*=1$, $\lambda=0.25$, $\mu=0.15$, $T=0.25$, an $L=12$ triangular cluster, and $40$ positive and negative fermionic Matsubara frequencies. The momenta are $\mathbf q_1=(n_q/L)\mathbf b_1$, $\mathbf q_2=\mathbf q_3=0$, and $\mathbf q_0=-\mathbf q_1$, with ordered polarizations $(x,y,x)$. Including the full set of $6+12+3+4+1=26$ terms with the finite-$q$ Peierls vertex factors of Eq.~\eqref{eqpeierls-line-supp} cancels the residual to numerical roundoff, confirming that contact terms are required before the $q$-linear magnetization projection.
}
\label{figward-check-supp}
\end{figure}

The compact vertices in Eqs.~\eqref{eqGamma1-supp} through
\eqref{eqGamma4-supp} are the uniform-field limits of their finite-momentum
counterparts. In a Peierls lattice regularization the wavevector also enters
the vertices directly. For a hopping matrix $T_{\mathbf d}$ on a straight
bond from $\mathbf R$ to $\mathbf R+\mathbf d$, a source leg with wavevector
$\mathbf q_r$ and polarization $\mathbf e_r$ contributes the line integral
\begin{equation}
\Phi_r(\mathbf R,\mathbf d)=
(\mathbf e_r\cdot\mathbf d)
\exp\!\left[i\mathbf q_r\cdot\left(\mathbf R+\frac{\mathbf d}{2}\right)\right]
\operatorname{sinc}\!\left(\frac{\mathbf q_r\cdot\mathbf d}{2}\right),
\qquad \operatorname{sinc}x\equiv\frac{\sin x}{x}.
\label{eqpeierls-line-supp}
\end{equation}
The $n$-leg Peierls vertex follows from the product of the $n$ corresponding
line-integral factors and the Peierls phase differentiated to order $n$. At
finite lattice spacing the $q$ derivative in the magnetization projection
then acts on both the propagator routing and these vertex form factors, so
the Ward-complete kernel retains both contributions. Figure~\ref{figward-check-supp} gives a numerical version of the Ward-identity
cancellation in Eq.~\eqref{eqward-output-supp}. Incomplete kernels contain
spurious longitudinal pieces, whereas the full 26-term kernel is transverse
within numerical precision.

Setting $A=0$, replacing $X\to\Gamma$, and assigning the external bosonic four-momenta $Q_r\equiv(\bq_r,\ii\omega_{rm})$ to the three incoming fields gives the ordered Matsubara kernel, with $Q=Q_1+Q_2+Q_3$ and $\beta_T\equiv 1/T$
\begin{equation}
\wt K_{a,jkl}(Q,Q_1,Q_2,Q_3)=\frac{1}{\beta_T V}\sum_K\Tr\Big[\wt{\cal B}_{a,jkl}(K)+\wt{\cal C}_{a,jkl}(K)+\wt{\cal D}_{a,jkl}(K)+\wt{\cal Q}_{a,jkl}(K)\Big].
\label{eqKthird-supp}
\end{equation}
The objects $\wt{\cal B}$, $\wt{\cal C}$, $\wt{\cal D}$, and $\wt{\cal Q}$ are the momentum-space versions of Eqs.~\eqref{eqBclass-supp} through \eqref{eqQclass-supp} after the replacement $X\to\Gamma$, with the internal Green-function arguments determined by momentum conservation along the matter loop. Defining the partial sums $P_0\equiv 0$, $P_1\equiv Q_{\pi_1}$, $P_2\equiv Q_{\pi_1}+Q_{\pi_2}$, $P_3\equiv Q_1+Q_2+Q_3=Q$, the explicit forms read
\begin{align}
\wt{\calB}_{a,jkl}(K)
&=-\sum_{\pi\in S_3}
G_{K+P_3}\Gamma^{(1)}_{\pi_3}\,G_{K+P_2}\Gamma^{(1)}_{\pi_2}\,G_{K+P_1}\Gamma^{(1)}_{\pi_1}\,G_K\Gamma^{(1)}_a,
\label{eqBtilde-supp}\\
\wt{\calC}_{a,jkl}(K)
&=\Big(G_{K+Q}\Gamma^{(2)}_{jk}\,G_{K+Q_3}\Gamma^{(1)}_l\,G_K\Gamma^{(1)}_a
+G_{K+Q}\Gamma^{(1)}_l\,G_{K+Q_1+Q_2}\Gamma^{(2)}_{jk}\,G_K\Gamma^{(1)}_a
\nonumber\\
&\quad+G_{K+Q}\Gamma^{(2)}_{jl}\,G_{K+Q_2}\Gamma^{(1)}_k\,G_K\Gamma^{(1)}_a
+G_{K+Q}\Gamma^{(1)}_k\,G_{K+Q_1+Q_3}\Gamma^{(2)}_{jl}\,G_K\Gamma^{(1)}_a
\nonumber\\
&\quad+G_{K+Q}\Gamma^{(2)}_{kl}\,G_{K+Q_1}\Gamma^{(1)}_j\,G_K\Gamma^{(1)}_a
+G_{K+Q}\Gamma^{(1)}_j\,G_{K+Q_2+Q_3}\Gamma^{(2)}_{kl}\,G_K\Gamma^{(1)}_a\Big)
\nonumber\\
&\quad+\Big(G_{K+Q}\Gamma^{(1)}_j\,G_{K+Q_2+Q_3}\Gamma^{(1)}_k\,G_K\Gamma^{(2)}_{al}
+G_{K+Q}\Gamma^{(1)}_k\,G_{K+Q_1+Q_3}\Gamma^{(1)}_j\,G_K\Gamma^{(2)}_{al}
\nonumber\\
&\quad+G_{K+Q}\Gamma^{(1)}_j\,G_{K+Q_2+Q_3}\Gamma^{(1)}_l\,G_K\Gamma^{(2)}_{ak}
+G_{K+Q}\Gamma^{(1)}_l\,G_{K+Q_1+Q_2}\Gamma^{(1)}_j\,G_K\Gamma^{(2)}_{ak}
\nonumber\\
&\quad+G_{K+Q}\Gamma^{(1)}_k\,G_{K+Q_1+Q_3}\Gamma^{(1)}_l\,G_K\Gamma^{(2)}_{aj}
+G_{K+Q}\Gamma^{(1)}_l\,G_{K+Q_1+Q_2}\Gamma^{(1)}_k\,G_K\Gamma^{(2)}_{aj}\Big),
\label{eqCtilde-supp}\\
\wt{\calD}_{a,jkl}(K)
&=-\Big(G_{K+Q}\Gamma^{(2)}_{jk}\,G_K\Gamma^{(2)}_{al}
+G_{K+Q}\Gamma^{(2)}_{jl}\,G_K\Gamma^{(2)}_{ak}
+G_{K+Q}\Gamma^{(2)}_{kl}\,G_K\Gamma^{(2)}_{aj}\Big)
\nonumber\\
&\quad-\Big(G_{K+Q}\Gamma^{(3)}_{jkl}\,G_K\Gamma^{(1)}_a
+G_{K+Q}\Gamma^{(1)}_j\,G_K\Gamma^{(3)}_{akl}
\nonumber\\
&\hspace{6em}+G_{K+Q}\Gamma^{(1)}_k\,G_K\Gamma^{(3)}_{ajl}
+G_{K+Q}\Gamma^{(1)}_l\,G_K\Gamma^{(3)}_{ajk}\Big),
\label{eqDtilde-supp}\\
\wt{\calQ}_{a,jkl}(K)
&=G_K\Gamma^{(4)}_{ajkl}.
\label{eqQtilde-supp}
\end{align}
The momentum routings follow from the convention that $A_a(-Q)$ enters the response-leg vertex and from momentum conservation at every vertex. Thus each term in Eqs.~\eqref{eqBtilde-supp} through \eqref{eqQtilde-supp} is obtained by replacing the operator sequence in Eqs.~\eqref{eqBclass-supp} through \eqref{eqQclass-supp} by the corresponding partial sums of the incoming four-momenta along the loop.

\section{Retarded kernel and practical \texorpdfstring{$q\to0$}{q -> 0} extraction}
\label{secmagnetization-extraction}

The ordered Matsubara kernel in Eq.~\eqref{eqKthird-supp} is continued to the ordered retarded response only after the causal ordering of the paired input labels is fixed. For the ordering $(j,\omega_1),(k,\omega_2),(l,\omega_3)$ used below, the intraband denominators depend on the tail frequency sums. The fully retarded branch is represented by
\begin{equation}
\ii\omega_{3m}\to\omega_3+\ii0^+,
\qquad
\ii(\omega_{2m}+\omega_{3m})\to\omega_2+\omega_3+\ii0^+,
\qquad
\ii\Omega_m\to\Omega+\ii0^+.
\label{eqanalytic-cont-supp}
\end{equation}
Here $\Omega_m=\omega_{1m}+\omega_{2m}+\omega_{3m}$ and $\Omega=\omega_1+\omega_2+\omega_3$. This is the nested commutator, or equivalently Keldysh, causal branch for the chosen ordered labels. It should not be read as three independent leg continuations. Other causal branches are obtained by permuting the paired labels before applying the same rule and then using the symmetrization of Eq.~\eqref{eqbeta-phys-sym-supp}.

To extract the cubic magnetization tensor we use the fact that the bound (magnetization) part of the current is related to the magnetization density $M_c$ by Amp\`ere's law, $\mathbf j^{(M)}=\bm\nabla\times\mathbf M$, which in Fourier space at small $\bq$ reads
\begin{equation}
j_a^{(M)}(\bq)=\ii\,\epsilon_{abc}\,q_b\,M_c(\bq)+\order{q^2}.
\label{eqjM-curl-supp}
\end{equation}
Antisymmetrizing in the pair $(a,b)$ (factor 2) and using the curl identity $\epsilon_{cab}\epsilon_{abd}=2\delta_{cd}$ (another factor 2) inverts Eq.~\eqref{eqjM-curl-supp},
\begin{equation}
M_c(\bq=0)
=\frac{1}{4\,\ii}\,\epsilon_{cab}\!\left[\frac{\partial j_a^{(M)}}{\partial q_b}-\frac{\partial j_b^{(M)}}{\partial q_a}\right]_{\bq=0}.
\label{eqM-from-j-supp}
\end{equation}
This identifies the magnetization with the antisymmetric, $q$-linear part of the current, as in the modern theory of orbital magnetization~\cite{shi2007orbital,xiao2010berry,thonhauser2005orbital,ceresoli2006orbital}. To convert from the vector-potential kernel $\wt K^R_{a,jkl}$, which couples to $A_jA_kA_l$, to a kernel that couples to $E_jE_kE_l$, we use $\mathbf E_r=-\partial_t\mathbf A_r$, i.e.\ $E_r(Q_r)=\ii\omega_r A_r(Q_r)$ at $\bq_r=0$. Hence
\begin{equation}
j_a^{(3)}(Q)=\wt K^R_{a,jkl}(Q,Q_1,Q_2,Q_3)\,A_j A_k A_l
=\frac{\wt K^R_{a,jkl}}{(\ii\omega_1)(\ii\omega_2)(\ii\omega_3)}\,E_jE_kE_l
=\frac{\ii\,\wt K^R_{a,jkl}}{\omega_1\omega_2\omega_3}\,E_jE_kE_l.
\label{eqjE-from-jA-supp}
\end{equation}
Identifying the magnetization-current part of $j_a^{(3)}$ in Eq.~\eqref{eqM-from-j-supp} and writing $M_c=\wt\beta_{cjkl}E_jE_kE_l$ then gives the cubic magnetization-tensor formula
\begin{equation}
\wt\beta_{cjkl}(\Omega,\omega_1,\omega_2,\omega_3)
=
\frac{1}{4\,\ii}\,\epsilon_{cab}
\left[\frac{\partial}{\partial q_b}\frac{\ii\,\wt K^R_{a,jkl}}{\omega_1\omega_2\omega_3}-\frac{\partial}{\partial q_a}\frac{\ii\,\wt K^R_{b,jkl}}{\omega_1\omega_2\omega_3}\right]_{\bq=0}
=
\frac{1}{4\omega_1\omega_2\omega_3}\epsilon_{cab}
\left[\frac{\partial\wt K^R_{a,jkl}}{\partial q_b}-\frac{\partial\wt K^R_{b,jkl}}{\partial q_a}\right]_{\bq=0},
\label{eqexact-kubo-supp}
\end{equation}
where $\epsilon_{cab}$ is the Levi-Civita symbol and the derivative is taken with respect to the total output wavevector $\bq$ at $\bq=0$. The prefactor $1/4$ combines the factor of two from antisymmetrizing in $(a,b)$ with the factor of two from the curl identity, the factor $1/(\omega_1\omega_2\omega_3)$ converts the vector-potential kernel to the electric-field kernel through Eq.~\eqref{eqjE-from-jA-supp}. The antisymmetric $q$-linear projection isolates the magnetization-current part of the response, consistent with modern orbital-magnetization formulations and recent links between nonequilibrium orbital magnetization and nonlinear Hall transport~\cite{shi2007orbital,xiao2010berry,zhang2025orbital}.

The ordered tensor in Eq.~\eqref{eqexact-kubo-supp} follows the convention fixed in Eq.~\eqref{eqbeta-phys-sym-supp}, the point-group tables below use $\beta^{\mathrm{phys}}$.

The derivative in Eq.~\eqref{eqexact-kubo-supp} should be understood with momentum conservation enforced. In a translationally invariant kernel the four external momenta satisfy $Q_0+Q_1+Q_2+Q_3=0$, where $Q_0$ is the source momentum on the measured current leg. The small vector $\bq$ can therefore be routed through any one of the incoming fields, distributed among the three incoming fields as a slow envelope momentum of the product $E_jE_kE_l$, or equivalently introduced as a magnetic probe coupled to the composite current response. These prescriptions are equivalent in the $q\to0$ antisymmetric projection after the Ward identities are imposed. In the finite-difference notation below, $\wt K^R_{a,jkl}(q\hat{\mathbf b})$ is a shorthand for this momentum-conserving long-wavelength limit, for example $\bq_1=q\hat{\mathbf b}$, $\bq_2=\bq_3=0$, and $\bq_0=-\bq_1$, followed by the input-leg symmetrization of Eq.~\eqref{eqbeta-phys-sym-supp} when the physical tensor is desired. Approximating the derivatives in Eq.~\eqref{eqexact-kubo-supp} by symmetric finite differences along $\hat{\mathbf a}$ and $\hat{\mathbf b}$ gives
\begin{equation}
\wt\beta_{cjkl}
=
\frac{1}{4\omega_1\omega_2\omega_3}\epsilon_{cab}\lim_{q\to0}
\frac{\wt K^R_{a,jkl}(q\hat{\mathbf b})-\wt K^R_{a,jkl}(-q\hat{\mathbf b})-\wt K^R_{b,jkl}(q\hat{\mathbf a})+\wt K^R_{b,jkl}(-q\hat{\mathbf a})}{2q}.
\label{eqcentered-diff-supp}
\end{equation}
Here $\hat{\mathbf a}$ and $\hat{\mathbf b}$ are unit vectors along the Cartesian $a$ and $b$ directions. Only the magnitude $q\to0$ enters the limit, and the four kernel evaluations along $\pm\hat{\mathbf a}$ and $\pm\hat{\mathbf b}$ implement the antisymmetric $q$-linear projection in Eq.~\eqref{eqM-from-j-supp}. Dropping contact terms can spoil the Ward-identity cancellations and introduce spurious gauge or routing dependence in the extracted magnetization, as in incomplete gauge-invariant reorganizations of nonlinear optical responses~\cite{aversa1995nonlinear,parker2019diagrammatic}.

\section{Gauge-covariant derivation of the static cubic moment tensor}
\label{secstatic-cubic-moment}
We construct the static second-order correction to the orbital moment in a
gauge-covariant form, starting from the mixed electric-magnetic component of
the second-order positional-shift functional~\cite{gao2014field} and
following the semiclassical treatment of wave-packet geometry, orbital
magnetization, and field-corrected band
energies~\cite{sundaram1999wavepacket,shi2007orbital,xiao2010berry,gao2015geometrical}.
Throughout, $\mu,\nu,\eta$ are band indices, $i,j,k,\alpha,\beta,\gamma$ are
Cartesian directions, and $H\equiv h(\bk)$ is the Bloch Hamiltonian with
cell-periodic eigenstates $|\mu\rangle\equiv|u_\mu(\bk)\rangle$ obeying
$h(\bk)|u_\mu\rangle=\eps_\mu|u_\mu\rangle$ and
$\langle u_\mu|u_\nu\rangle=\delta_{\mu\nu}$, where $\eps_\mu(\bk)$ is the
band energy. Define the first-order interband mixing amplitudes
\begin{equation}
R^j_{\mu\nu}(\bk)=\frac{A^j_{\mu\nu}(\bk)}{\Delta_{\nu\mu}(\bk)}
=\ii\frac{\langle \mu|\partial_jH|\nu\rangle}{\Delta_{\nu\mu}^2},
\qquad
C^k_{\mu\nu}(\bk)=-\frac{m^k_{\mu\nu}(\bk)}{\Delta_{\nu\mu}(\bk)},
\label{eqRC-def-supp}
\end{equation}
where $A^j_{\mu\nu}=\langle\mu|\ii\partial_j|\nu\rangle$ is the interband
Berry connection, $\Delta_{\nu\mu}=\eps_\nu-\eps_\mu$ the band-energy
difference, and $m^k_{\mu\nu}$ the interband orbital-moment matrix element.
The derivatives $\partial_j$ act on the cell-periodic states in a smooth
gauge, so that $\langle\mu|\partial_jH|\nu\rangle=-\ii\Delta_{\nu\mu}A^j_{\mu\nu}$
for $\mu\neq\nu$, which gives the second equality in
Eq.~\eqref{eqRC-def-supp}. The orbital-moment matrix element is the
off-diagonal generalization of the single-band
moment~\cite{xiao2010berry,shi2007orbital,thonhauser2005orbital,ceresoli2006orbital,sundaram1999wavepacket},
\[
m^k_{\mu\nu}(\bk)
=
\frac{\ii e}{2\hbar}\,\epsilon_{kab}\,
\langle \partial_a u_\mu|\,\big[h(\bk)-\tfrac12(\eps_\mu+\eps_\nu)\big]\,|\partial_b u_\nu\rangle,
\]
with $\epsilon_{kab}$ the Levi-Civita symbol; for $\mu=\nu$ it reduces to the
self-rotation moment $m_k^{\nu\nu}\equiv m_k^\nu$. The gauge-covariant
derivative of an interband object $\mathcal{Y}_{\mu\nu}$ is
\begin{equation}
D^{\mu\nu}_{\alpha}\mathcal{Y}_{\mu\nu}=\left(A^{\nu}_{\alpha}-A^{\mu}_{\alpha}-\ii\partial_{\alpha}\right)\mathcal{Y}_{\mu\nu},
\label{eqcov-deriv-supp}
\end{equation}
with $A^{\mu}_{\alpha}=\langle\mu|\ii\partial_{\alpha}|\mu\rangle$ the
intraband Berry connection; under $|\mu\rangle\to e^{\ii\varphi_\mu}|\mu\rangle$
the connection terms cancel the phase gradient of $\partial_\alpha \mathcal{Y}_{\mu\nu}$,
so that $D^{\mu\nu}_\alpha \mathcal{Y}_{\mu\nu}\to e^{\ii(\varphi_\nu-\varphi_\mu)}D^{\mu\nu}_\alpha \mathcal{Y}_{\mu\nu}$.
This is the length-gauge generalized derivative of Aversa and
Sipe~\cite{aversa1995nonlinear,sipe2000nonlinear,parker2019diagrammatic}
adapted to the positional-shift formalism~\cite{gao2014field,gao2015geometrical}.

For generic interband amplitudes $\mathcal{X}$ and $\mathcal{Y}$, each
standing for either $R$ or $C$ of Eq.~\eqref{eqRC-def-supp}, define the two
bilinear blocks
\begin{align}
U^{\nu\mu}_{\alpha\beta\gamma}[\mathcal{X},\mathcal{Y}]&=\mathcal{X}^{\beta}_{\nu\mu}D^{\mu\nu}_{\alpha}\mathcal{Y}^{\gamma}_{\mu\nu},
\label{eqU-supp}\\
V^{\nu\mu\eta}_{\alpha\beta\gamma}[\mathcal{X},\mathcal{Y}]&=\left(2\mathcal{X}^{\alpha}_{\nu\mu}A^{\beta}_{\mu\eta}+\mathcal{X}^{\beta}_{\nu\mu}A^{\alpha}_{\mu\eta}\right)\mathcal{Y}^{\gamma}_{\eta\nu},
\label{eqV-supp}
\end{align}
which appear after the second-order positional shift is reorganized
covariantly. The $U$ block carries the covariant derivative of a two-band
amplitude, and the $V$ block carries a genuine intermediate-band insertion
summed over $\eta\neq\mu,\nu$. The asymmetric coefficient $(2,1)$ in
Eq.~\eqref{eqV-supp} is fixed by the pure-electric limit. Setting
$\mathcal{X}=\mathcal{Y}=R$,
\begin{equation}
V^{\nu\mu\eta}_{\alpha\beta\gamma}[R,R]
=\left(2R^\alpha_{\nu\mu}A^\beta_{\mu\eta}+R^\beta_{\nu\mu}A^\alpha_{\mu\eta}\right)R^\gamma_{\eta\nu},
\label{eqV-pure-electric-supp}
\end{equation}
and substitution into Eq.~\eqref{eqTfunctional-supp} reproduces the
Gao-Yang-Niu positional shift only when the coefficient of
$R^\alpha_{\nu\mu}A^\beta_{\mu\eta}R^\gamma_{\eta\nu}$ is twice that of
$R^\beta_{\nu\mu}A^\alpha_{\mu\eta}R^\gamma_{\eta\nu}$, that is
\begin{equation}
c_{\alpha\beta}=2,\qquad c_{\beta\alpha}=1.
\label{eqV-two-one-supp}
\end{equation}

The second-order positional-shift functional is quadratic in the first-order
amplitudes and, in covariant form, reduces to the blocks of
Eqs.~\eqref{eqU-supp} and \eqref{eqV-supp}. For a generic covariant amplitude
$Z$ (again $R$ or $C$), with both block arguments set equal,
\begin{align}
T^{\nu}_{\alpha\beta\gamma}[Z]
=\mathrm{Re}\sum_{\mu\neq\nu}\Bigg[
U^{\nu\mu}_{\alpha\beta\gamma}[Z,Z]+U^{\nu\mu}_{\beta\alpha\gamma}[Z,Z]-U^{\mu\nu}_{\beta\gamma\alpha}[Z,Z]-\sum_{\eta\neq\mu,\nu}V^{\nu\mu\eta}_{\alpha\beta\gamma}[Z,Z]\Bigg].
\label{eqTfunctional-supp}
\end{align}
The first two $U$ terms symmetrize $(\alpha,\beta)$, and the remaining $U$
and $V$ terms reproduce the pure-electric positional shift when $Z=R$. To
isolate the mixed electric-magnetic part, set $Z=sR+tC$ with bookkeeping
parameters $s,t$. Since every block is bilinear,
\begin{equation}
T^\nu_{\alpha\beta\gamma}[sR+tC]
=s^2T^\nu_{\alpha\beta\gamma}[R]+t^2T^\nu_{\alpha\beta\gamma}[C]+st\,\calQ^\nu_{\alpha\beta\gamma},
\label{eqT-st-expansion-supp}
\end{equation}
and $\partial_s\partial_t$ at $s=t=0$ selects the two ordered cross terms,
\begin{align}
\calQ^{\nu}_{\alpha\beta\gamma}
&=\left.\frac{\partial^2}{\partial s\,\partial t}T^{\nu}_{\alpha\beta\gamma}[sR+tC]\right|_{s=t=0}
\nonumber\\
&=\mathrm{Re}\sum_{\mu\neq\nu}\Big[U^{\nu\mu}_{\alpha\beta\gamma}[R,C]+U^{\nu\mu}_{\alpha\beta\gamma}[C,R]+U^{\nu\mu}_{\beta\alpha\gamma}[R,C]+U^{\nu\mu}_{\beta\alpha\gamma}[C,R]
\nonumber\\
&\qquad\qquad\qquad-U^{\mu\nu}_{\beta\gamma\alpha}[R,C]-U^{\mu\nu}_{\beta\gamma\alpha}[C,R]
\nonumber\\
&\qquad\qquad\qquad-\sum_{\eta\neq\mu,\nu}\big(V^{\nu\mu\eta}_{\alpha\beta\gamma}[R,C]+V^{\nu\mu\eta}_{\alpha\beta\gamma}[C,R]\big)\Big].
\label{eqQcross-supp}
\end{align}
Since $R$ is linear in the electric field and $C$ in the magnetic field,
$\calQ^\nu_{\alpha\beta\gamma}$ is the mixed $EB$ contribution and yields the
gauge-invariant second-order positional shift
\begin{equation}
a_{\alpha}^{(2,EB),\nu}(\bk)=\calQ^{\nu}_{\alpha\beta\gamma}(\bk)E_{\beta}B_{\gamma},
\label{eqa2EB-supp}
\end{equation}
linear in both the electric field $E_\beta$ and the magnetic field $B_\gamma$.
To relate $\calQ$ to the cubic moment, the electric field entering
Eq.~\eqref{eqa2EB-supp} in the perturbative normalization is the electric
force $\mathcal E_j=eE_j$, and the mixed shift contributes the $\mathcal E^2B$
energy $+\tfrac12\mathcal E_j\mathcal E_kB_i(\calQ^\nu_{jki}+\calQ^\nu_{kji})$,
the factor $1/2$ symmetrizing the two identical electric fields. Matching to
the Zeeman form $-B_i\delta m^{(2)}_{\nu,i}$ with
$\delta m^{(2)}_{\nu,i}=e^2H^\nu_{ijk}E_jE_k$ gives
\begin{equation}
H^{\nu}_{ijk}=-\frac{1}{2}\left(\calQ^{\nu}_{jki}+\calQ^{\nu}_{kji}\right).
\label{eqH-explicit-supp}
\end{equation}
This $1/2$ is the field-symmetrization factor for two identical electric
fields, distinct from the $1/4$ in Eq.~\eqref{eqM-from-j-supp}, which arose
from antisymmetrization and the curl identity. The charge and symmetrization
factors are therefore fixed by the chain
\begin{align}
a_{\alpha}^{(2,EB),\nu}&=\calQ^\nu_{\alpha\beta\gamma}E_\beta B_\gamma,
&
H^\nu_{ijk}&=-\frac12(\calQ^\nu_{jki}+\calQ^\nu_{kji}),\nonumber\\
\delta m^{(2)}_{\nu,i}&=e^2H^\nu_{ijk}E_jE_k,
&
\beta^{(H)}_{ijkl}&=-e^3\tau\sum_\nu\int_{\rm BZ}\frac{\dd^dk}{(2\pi)^d}H^\nu_{ijk}v_l^\nu f_0'(\eps_\nu),
\label{eqH-charge-counting-supp}
\end{align}
where $v_l^\nu=\hbar^{-1}\partial_l\eps_\nu$ is the band velocity, $f_0$ is the
equilibrium Fermi function, and $\tau$ is the relaxation time.

\section{Ordered occupation dynamics and full low-frequency formulas}
\label{seclowfreq}

We next collect the ordered low-frequency Boltzmann contributions. This
construction is logically distinct from the diagrammatic Kubo route of
Secs.~\ref{seckubo-kernel}-\ref{seckernel-reduction}; it builds the
magnetization from a wave-packet kinetic chain with a \emph{field-dressed}
local orbital moment $\wt m_{\nu,i}$ and recovers the same three dc
contributions. The equivalence holds only when the second-order
positional-shift correction $\delta m^{(2)}$ entering the mixed-quadrupole
contribution and the metric-induced energy shift $\delta\eps^{(2)}$ entering
the metric contribution are retained explicitly. Let $f_0(\eps_\nu)$ be the
equilibrium Fermi function, $f_0'\equiv\partial_{\eps}f_0$, and
$v_l^{\nu}(\bk)\equiv(1/\hbar)\partial_l\eps_\nu(\bk)$ the band velocity, and
let $\delta\wt f^{(n)}_{\nu,r\cdots3}$ denote the $n$th-order occupation
correction of band $\nu$ for a fixed field ordering. The ordered Boltzmann
chain is generated recursively by the standard semiclassical logic of
nonlinear Hall and magneto-optical
transport~\cite{sodemann2015quantum,xiao2019theory,nandy2019symmetry,morimoto2016semiclassical}
\begin{equation}
\left[-\ii(\omega_r+\cdots+\omega_3)+\tau^{-1}\right]\delta\wt f^{(n)}_{\nu,r\cdots3}
=-\frac{e}{\hbar}E_r\partial_r\,\delta\wt f^{(n-1)}_{\nu,(r+1)\cdots3},
\label{eqBoltzmann-recursion-supp}
\end{equation}
with $\delta\wt f^{(0)}_{\nu}=f_0(\eps_\nu)$ and
$\partial_r\equiv\partial/\partial k_r$, where the label $r$ marks the next
field insertion in the ordered chain. For the ordered triple
$(j,\omega_1),(k,\omega_2),(l,\omega_3)$ we obtain
\begin{align}
\delta\wt f^{(1)}_{\nu,l}
&=-\frac{e\tau}{1-\ii\omega_3\tau}E_lv_l^{\nu}f_0'(\eps_\nu),
\label{eqf1-supp}\\
\delta\wt f^{(2)}_{\nu,kl}
&=\frac{e^2\tau^2}{\hbar}E_kE_l\frac{\partial_k\left[v_l^{\nu}f_0'(\eps_\nu)\right]}{[1-\ii(\omega_2+\omega_3)\tau](1-\ii\omega_3\tau)},
\label{eqf2-supp}\\
\delta\wt f^{(3)}_{\nu,jkl}
&=-\frac{e^3\tau^3}{\hbar^2}E_jE_kE_l\frac{\partial_j\partial_k\left[v_l^{\nu}f_0'(\eps_\nu)\right]}{(1-\ii\Omega\tau)[1-\ii(\omega_2+\omega_3)\tau](1-\ii\omega_3\tau)}.
\label{eqf3-supp}
\end{align}
The electric-field part of the second-order wave-packet energy is controlled
by the gap-weighted (band-normalized) metric
\begin{equation}
\mathcal{G}^{\nu}_{jk}(\bk)
=
2\,\mathrm{Re}\sum_{\mu\neq\nu}\frac{A^{j}_{\nu\mu}(\bk)A^{k}_{\mu\nu}(\bk)}{\Delta_{\nu\mu}(\bk)}
=
2\,\mathrm{Re}\sum_{\mu\neq\nu}\frac{\langle \nu|\partial_jH|\mu\rangle\langle \mu|\partial_kH|\nu\rangle}{\Delta_{\nu\mu}^3(\bk)},
\label{eqGmetric-def-supp}
\end{equation}
a symmetric tensor that follows from nondegenerate second-order perturbation
theory and controls the static electric polarizability of band
$\nu$~\cite{gao2014field,gao2015geometrical}. Thus
\begin{equation}
\delta\eps^{(2)}_{\nu}(\bk)=\frac{e^2}{2}\mathcal{G}^{\nu}_{jk}(\bk)E_jE_k,
\label{eqeps2-supp}
\end{equation}
which acts as an additional source term in the ordered Boltzmann equation and
gives
\begin{equation}
\delta\wt f^{(G)}_{\nu,jkl}
=-\frac{e^3\tau}{2\hbar}E_jE_kE_l\frac{\partial_l\left[\mathcal{G}^{\nu}_{jk}(\bk)f_0'(\eps_\nu)\right]}{1-\ii\Omega\tau},
\label{eqfG-supp}
\end{equation}
the label $(G)$ marking the occupation correction sourced by the metric
energy shift. We combine this with the moment expansion
\begin{equation}
\wt m_{\nu,i}=m_i^{\nu}+\delta m^{(1)}_{\nu,i}+\delta m^{(2)}_{\nu,i}+\cdots,
\qquad
M_i=\frac{1}{V}\sum_{\nu,\bk}\wt m_{\nu,i}f_\nu,
\label{eqmoment-expansion-supp}
\end{equation}
where $\wt m_{\nu,i}$ is the field-dressed local orbital moment, $f_\nu$ the
occupation of band $\nu$, and $M_i$ the magnetization density. The
equilibrium Berry-curvature contribution to $M_i$ does not generate an
independent cubic dc term; after integration by parts it is absorbed into the
field-dressed moment and band energy of the mixed-quadrupole and metric
contributions.

For the ordered triple $(j,\omega_1),(k,\omega_2),(l,\omega_3)$, define the
ordered moment kernels
$$
\delta m^{(1)}_{\nu,i,j}(\bk,\omega_1)=\mathcal M^{\nu}_{ij}(\bk,\omega_1)E_j,
\qquad
\delta m^{(2)}_{\nu,i,jk}(\bk,\omega_1,\omega_2)=\mathcal N^{\nu}_{ijk}(\bk,\omega_1,\omega_2)E_jE_k.
$$
At the adiabatic order retained, the first-order electric positional shift is
$a_{a}^{(1,E),\nu}(\bk)=e\,\mathcal{G}^{\nu}_{aj}(\bk)E_j$, consistent with the
charge counting in Eq.~\eqref{eqeps2-supp}, and the induced first-order moment
is
$$
\delta m^{(1)}_{\nu,i}
=-\frac{e}{2}\,\epsilon_{iab}\,a_{a}^{(1,E),\nu}(\bk)\,v_b^{\nu}(\bk)
=\mathcal M^{\nu}_{ij}(\bk,0)E_j,
\qquad
\mathcal M^{\nu}_{ij}(\bk,0)=-\frac{e^2}{2}\,\epsilon_{iab}\,\mathcal{G}^{\nu}_{aj}(\bk)\,v_b^{\nu}(\bk).
$$
This $\mathcal M\times\delta f^{(2)}$ term is retained in
Eq.~\eqref{eqbeta-low-general-supp} for a generic noncentrosymmetric magnetic
crystal. The mixed quadrupolar shift of Eq.~\eqref{eqH-explicit-supp} gives
$$
\mathcal N^{\nu}_{ijk}(\bk,0,0)=e^2 H^{\nu}_{ijk}(\bk).
$$
The tensor below uses the leading static
limits $\mathcal M^{\nu}_{ij}(\bk,\omega_1)=\mathcal M^{\nu}_{ij}(\bk,0)+\mathcal O(\omega_1)$
and $\mathcal N^{\nu}_{ijk}(\bk,\omega_1,\omega_2)=e^2 H^{\nu}_{ijk}(\bk)+\mathcal O(\omega_1,\omega_2)$.
Keeping only terms third order in the fields,
\[
\delta\wt M^{(3)}_{i,jkl}
=
\frac{1}{V}\sum_{\nu,\bk}\Big[
\delta m^{(2)}_{\nu,i,jk}\,\delta\wt f^{(1)}_{\nu,l}
+\delta m^{(1)}_{\nu,i,j}\,\delta\wt f^{(2)}_{\nu,kl}
+m_i^{\nu}\,\delta\wt f^{(G)}_{\nu,jkl}
+m_i^{\nu}\,\delta\wt f^{(3)}_{\nu,jkl}
\Big],
\]
and substituting Eqs.~\eqref{eqf1-supp}, \eqref{eqf2-supp}, \eqref{eqf3-supp},
and \eqref{eqfG-supp} and removing the common factor $E_jE_kE_l$ gives
\begin{align}
\wt\beta^{\mathrm{low}}_{ijkl}(\Omega,\omega_1,\omega_2,\omega_3)
&=-\frac{e\tau}{V}\sum_{\nu,\bk}\frac{\mathcal N^{\nu}_{ijk}(\bk,\omega_1,\omega_2)\,v_l^{\nu}(\bk)f_0'(\eps_\nu)}{1-\ii\omega_3\tau}
\nonumber\\
&\quad+\frac{e^2\tau^2}{\hbar V}\sum_{\nu,\bk}\frac{\mathcal M^{\nu}_{ij}(\bk,\omega_1)\,\partial_k\left[v_l^{\nu}(\bk)f_0'(\eps_\nu)\right]}{[1-\ii(\omega_2+\omega_3)\tau](1-\ii\omega_3\tau)}
\nonumber\\
&\quad-\frac{e^3\tau}{2\hbar V}\sum_{\nu,\bk}\frac{m_i^{\nu}(\bk)\,\partial_l\left[\mathcal{G}^{\nu}_{jk}(\bk)f_0'(\eps_\nu)\right]}{1-\ii\Omega\tau}
\nonumber\\
&\quad-\frac{e^3\tau^3}{\hbar^2V}\sum_{\nu,\bk}\frac{m_i^{\nu}(\bk)\,\partial_j\partial_k\left[v_l^{\nu}(\bk)f_0'(\eps_\nu)\right]}{(1-\ii\Omega\tau)[1-\ii(\omega_2+\omega_3)\tau](1-\ii\omega_3\tau)}.
\label{eqbeta-low-general-supp}
\end{align}
This decomposition is compatible with recent third-order transport
formulations that isolate distinct quantum-geometric
contributions~\cite{xiang2023third,mandal2024quantum,fang2024quantum}. In a
nonmagnetic time-reversal-symmetric crystal the second
($\mathcal M\,\delta f^{(2)}$) term is odd in $\bk$ and integrates to zero,
leaving the surviving ordered tensor
\begin{equation}
\wt\beta^{\mathrm{low,TR}}_{ijkl}(\Omega,\omega_1,\omega_2,\omega_3)
=\wt\beta^{\mathrm{low}}_{ijkl}(\Omega,\omega_1,\omega_2,\omega_3)\big|_{\mathcal M^{\nu}_{ij}\to0},
\label{eqbeta-low-TR-supp}
\end{equation}
i.e.\ the first, third, and fourth lines of Eq.~\eqref{eqbeta-low-general-supp}.
The denominators belong to the ordered insertion
$(j,\omega_1),(k,\omega_2),(l,\omega_3)$. The measured nondegenerate response
follows from the paired-label symmetrization in
Eq.~\eqref{eqbeta-phys-sym-supp}, which in the dc or degenerate THG limit
reduces to ordinary symmetrization over the last three Cartesian indices.

In the strict dc limit,
\begin{align}
\wt\beta^{\mathrm{dc,TR}}_{ijkl}
&=\frac{e^3\tau}{2V}\sum_{\nu,\bk}\left[\calQ^{\nu}_{jki}(\bk)+\calQ^{\nu}_{kji}(\bk)\right]v_l^{\nu}(\bk)f_0'(\eps_\nu)
\nonumber\\
&\quad-\frac{e^3\tau}{2\hbar V}\sum_{\nu,\bk}m_i^{\nu}(\bk)\Big[\big(\Gamma^{(\mathcal{G}),\nu}_{j,lk}+\Gamma^{(\mathcal{G}),\nu}_{k,lj}\big)f_0'(\eps_\nu)+\hbar \mathcal{G}^{\nu}_{jk}(\bk)v_l^{\nu}(\bk)f_0''(\eps_\nu)\Big]
\nonumber\\
&\quad-\frac{e^3\tau^3}{\hbar^2V}\sum_{\nu,\bk}m_i^{\nu}(\bk)\partial_j\partial_k\left[v_l^{\nu}(\bk)f_0'(\eps_\nu)\right],
\label{eqbeta-dc-TR-supp}
\end{align}
where
\begin{equation}
\Gamma^{(\mathcal{G}),\nu}_{c,ab}=\frac{1}{2}\left(\partial_a\mathcal{G}^{\nu}_{bc}+\partial_b\mathcal{G}^{\nu}_{ac}-\partial_c\mathcal{G}^{\nu}_{ab}\right)
\label{eqGammaG-supp}
\end{equation}
is the lowered-index Christoffel combination of the symmetric band-normalized
metric, satisfying
$\Gamma^{(\mathcal{G}),\nu}_{j,lk}+\Gamma^{(\mathcal{G}),\nu}_{k,lj}=\partial_l\mathcal{G}^{\nu}_{jk}$,
the identity used in Eq.~\eqref{eqbeta-dc-TR-supp}. For a two-band model,
$g^{\nu}_{jk}$ is the usual quantum metric and
$\Delta_{\nu\bar\nu}\equiv\eps_{\nu}-\eps_{\bar\nu}$ the signed separation to
the partner band $\bar\nu$,
\begin{equation}
\mathcal{G}^{\nu}_{jk}=\frac{2}{\Delta_{\nu\bar\nu}}g^{\nu}_{jk},
\label{eqG-g-two-band-supp}
\end{equation}
so that
\begin{equation}
\partial_l\mathcal{G}^{\nu}_{jk}=\frac{2}{\Delta_{\nu\bar\nu}}\left(\Gamma^{\nu}_{j,lk}+\Gamma^{\nu}_{k,lj}\right)-\frac{2\partial_l\Delta_{\nu\bar\nu}}{\Delta_{\nu\bar\nu}^2}g^{\nu}_{jk}.
\label{eqdG-Gamma-supp}
\end{equation}
with $\Gamma^{\nu}_{c,ab}$ the first-kind Christoffel symbols of $g^{\nu}$,
defined as in Eq.~\eqref{eqGammaG-supp} with $\mathcal{G}^{\nu}\to g^{\nu}$. The
metric-drift contribution can therefore be reorganized partly in Christoffel
form, with an additional gap-derivative term remaining. For related geometric
viewpoints on nonlinear response and Christoffel-type nonlinear magnetization,
see Refs.~\cite{ahn2022riemannian,qiang2026quantum,qian2026probing}.

\subsection{Integral and multipole form of \texorpdfstring{Eq.~\eqref{eqbeta-dc-TR-supp}}{the dc tensor}}
To expose the geometric structure of Eq.~\eqref{eqbeta-dc-TR-supp}, write
$$
\wt\beta^{\mathrm{dc,TR}}_{ijkl}=\beta^{(H)}_{ijkl}+\beta^{(G)}_{ijkl}+\beta^{(\mathrm{tr})}_{ijkl},
$$
where $\beta^{(H)}$ is the mixed-quadrupole contribution, $\beta^{(G)}$ the
metric or Christoffel contribution, and $\beta^{(\mathrm{tr})}$ the purely
transport contribution. The $f_0''$ term in Eq.~\eqref{eqbeta-dc-TR-supp}
comes from the chain rule
$\partial_l[\mathcal{G}^\nu_{jk}f_0'(\eps_\nu)]=(\partial_l\mathcal{G}^\nu_{jk})f_0'+\hbar\mathcal{G}^\nu_{jk}v_l^\nu f_0''$.
Using $V^{-1}\sum_{\bk}\to\int_{\rm BZ}\dd^dk/(2\pi)^d$ and integrating by
parts, the explicit Brillouin-zone integrals are
\begin{align}
\beta^{(H)}_{ijkl}
&=-e^3\tau\sum_{\nu}\int_{\rm BZ}\frac{\dd^dk}{(2\pi)^d}
H^{\nu}_{ijk}(\bk)\,v_l^{\nu}(\bk)\,f_0'\!\left(\eps_{\nu}\right)
\nonumber\\
&=\frac{e^3\tau}{\hbar}\sum_{\nu}\int_{\rm BZ}\frac{\dd^dk}{(2\pi)^d}
f_0\!\left(\eps_{\nu}\right)\,\partial_l H^{\nu}_{ijk}(\bk)
\nonumber\\
&=-\frac{e^3\tau}{2\hbar}\sum_{\nu}\int_{\rm BZ}\frac{\dd^dk}{(2\pi)^d}
f_0\!\left(\eps_{\nu}\right)\,\partial_l\!\left[\calQ^{\nu}_{jki}(\bk)+\calQ^{\nu}_{kji}(\bk)\right],
\label{eqbetaH-integral-supp}\\
\beta^{(G)}_{ijkl}
&=-\frac{e^3\tau}{2\hbar}\sum_{\nu}\int_{\rm BZ}\frac{\dd^dk}{(2\pi)^d}
m_i^{\nu}(\bk)\,\partial_l\!\left[\mathcal{G}^{\nu}_{jk}(\bk)f_0'\!\left(\eps_{\nu}\right)\right]
\nonumber\\
&=\frac{e^3\tau}{2\hbar}\sum_{\nu}\int_{\rm BZ}\frac{\dd^dk}{(2\pi)^d}
\left(\partial_l m_i^{\nu}(\bk)\right)\mathcal{G}^{\nu}_{jk}(\bk)f_0'\!\left(\eps_{\nu}\right),
\label{eqbetaG-integral-supp}\\
\beta^{(\mathrm{tr})}_{ijkl}
&=-\frac{e^3\tau^3}{\hbar^2}\sum_{\nu}\int_{\rm BZ}\frac{\dd^dk}{(2\pi)^d}
m_i^{\nu}(\bk)\,\partial_j\partial_k\!\left[v_l^{\nu}(\bk)f_0'\!\left(\eps_{\nu}\right)\right]
\nonumber\\
&=\frac{e^3\tau^3}{\hbar^3}\sum_{\nu}\int_{\rm BZ}\frac{\dd^dk}{(2\pi)^d}
f_0\!\left(\eps_{\nu}\right)\,\partial_j\partial_k\partial_l m_i^{\nu}(\bk).
\label{eqbetatr-integral-supp}
\end{align}
Equations~\eqref{eqbetaH-integral-supp}--\eqref{eqbetatr-integral-supp} agree
with the kernel-method forms Eqs.~\eqref{eqkr-betaH-intermediate-supp},
\eqref{eqkr-betaG-intermediate-supp}, and
\eqref{eqkr-betatr-intermediate-supp} derived in
Sec.~\ref{seckernel-reduction}, so the Boltzmann and diagrammatic routes give
the same three contributions. In the zero-temperature limit
$f_0'=-\delta(\eps_{\nu}-\mu)$, and Eq.~\eqref{eqbetaH-integral-supp} reduces
to the Fermi-surface form
\begin{equation}
\beta^{(H)}_{ijkl}=\frac{e^3\tau}{\hbar}\sum_{\nu}\int_{{\rm FS}_{\nu}}\frac{\dd S}{(2\pi)^d}
H^{\nu}_{ijk}(\bk)\,\hat n_l^{\nu}(\bk),
\label{eqbetaH-FS-supp}
\end{equation}
with $\hat n_l^{\nu}(\bk)=\partial_l\eps_{\nu}(\bk)/|\nabla_{\bk}\eps_{\nu}(\bk)|$.
Likewise,
\begin{equation}
\beta^{(G)}_{ijkl}=-\frac{e^3\tau}{2\hbar}\sum_{\nu}\int_{{\rm FS}_{\nu}}\frac{\dd S}{(2\pi)^d}
\frac{\left(\partial_lm_i^{\nu}\right)\mathcal{G}^{\nu}_{jk}}{|\nabla_{\bk}\eps_{\nu}|},
\label{eqbetaG-FS-supp}
\end{equation}
so the metric/Christoffel contribution is explicitly a Fermi-surface object
weighted by the gradient of the local orbital moment.

For the two-band reduction and for comparison with conventional
nonlinear-Hall language, introduce the axial Berry curvature and the band
orbital moment,
\begin{align}
\Omega_i^{\nu}(\bk)
&=\frac{1}{2}\epsilon_{iab}F_{ab}^{\nu}(\bk)
=-\epsilon_{iab}\,\mathrm{Im}\sum_{\mu\neq\nu}
\frac{\langle\nu|\partial_aH|\mu\rangle\langle\mu|\partial_bH|\nu\rangle}{\Delta_{\nu\mu}^2(\bk)},
\label{eqOmega-explicit-supp}\\
m_i^{\nu}(\bk)
&=-\frac{e}{2\hbar}\epsilon_{iab}\,\mathrm{Im}\sum_{\mu\neq\nu}
\frac{\langle\nu|\partial_aH|\mu\rangle\langle\mu|\partial_bH|\nu\rangle}{\Delta_{\nu\mu}(\bk)},
\label{eqm-explicit-supp}
\end{align}
with $F_{ab}^{\nu}$ the Berry curvature. Only two occupied-state multipoles
are needed for the dc formulas,
\begin{equation}
D^{(H)}_{l|ijk}
=\sum_{\nu}\int_{\rm BZ}\frac{\dd^dk}{(2\pi)^d}f_0(\eps_\nu)\partial_lH^\nu_{ijk},
\qquad
O^{(m)}_{jkli}
=\sum_{\nu}\int_{\rm BZ}\frac{\dd^dk}{(2\pi)^d}f_0(\eps_\nu)\partial_j\partial_k\partial_lm_i^\nu,
\label{eqneeded-multipoles-supp}
\end{equation}
so that Eqs.~\eqref{eqbetaH-integral-supp} and \eqref{eqbetatr-integral-supp}
become
\begin{equation}
\beta^{(H)}_{ijkl}=\frac{e^3\tau}{\hbar}D^{(H)}_{l|ijk},
\qquad
\beta^{(\mathrm{tr})}_{ijkl}=\frac{e^3\tau^3}{\hbar^3}O^{(m)}_{jkli}.
\label{eqbeta-multipole-shorthand-supp}
\end{equation}
The tensor $H^{\nu}_{ijk}$, or equivalently $\calQ^{\nu}_{\alpha\beta\gamma}$,
is a mixed electric and magnetic quadrupolar shift of the wave packet.

For the two-band reduction it is convenient to write the usual quantum metric
explicitly as
\begin{equation}
g^{\nu}_{jk}(\bk)
=\mathrm{Re}\sum_{\mu\neq\nu}A^{j}_{\nu\mu}(\bk)A^{k}_{\mu\nu}(\bk)
=\mathrm{Re}\sum_{\mu\neq\nu}
\frac{\langle\nu|\partial_jH|\mu\rangle\langle\mu|\partial_kH|\nu\rangle}{\Delta_{\nu\mu}^2(\bk)}.
\label{eqg-explicit-supp}
\end{equation}
This is the band-projected Fubini-Study quantum metric of Provost and
Vall\'ee~\cite{provost1980riemannian}, standard in the band-geometric and
quantum-information literature; see Refs.~\cite{xiao2010berry,resta2011insulating}
for reviews. With a single partner band $\bar\nu$,
Eqs.~\eqref{eqOmega-explicit-supp} and \eqref{eqm-explicit-supp} imply
\begin{equation}
m_i^{\nu}=\frac{e\,\Delta_{\nu\bar\nu}}{2\hbar}\Omega_i^{\nu},
\label{eqtwo-band-metric-moment-supp}
\end{equation}
which together with Eq.~\eqref{eqG-g-two-band-supp} closes the two-band
reduction. The metric contribution then becomes
\begin{align}
\beta^{(G)}_{ijkl}
&=\frac{e^4\tau}{2\hbar^2}\sum_{\nu}\int_{\rm BZ}\frac{\dd^dk}{(2\pi)^d}
g^{\nu}_{jk}(\bk)\Big[\partial_l\Omega_i^{\nu}(\bk)+\Omega_i^{\nu}(\bk)\partial_l\ln\Delta_{\nu\bar\nu}(\bk)\Big]f_0'\!\left(\eps_{\nu}\right),
\label{eqbetaG-berrydipole-supp}
\end{align}
a quantum-metric dressing of a Berry-curvature dipole plus a correction
controlled by the gradient of the interband gap, and the transport
contribution becomes
\begin{align}
\beta^{(\mathrm{tr})}_{ijkl}
&=\frac{e^4\tau^3}{2\hbar^4}\sum_{\nu}\int_{\rm BZ}\frac{\dd^dk}{(2\pi)^d}
f_0\!\left(\eps_{\nu}\right)\,\partial_j\partial_k\partial_l\!\left[\Delta_{\nu\bar\nu}(\bk)\Omega_i^{\nu}(\bk)\right].
\label{eqbetatr-berryoct-supp}
\end{align}
For a momentum-dependent gap, expanding the derivative in
Eq.~\eqref{eqbetatr-berryoct-supp} generates gap-gradient corrections to a
pure Berry-curvature octupole, so the compact form is more transparent than
the full derivative hierarchy. The mixed-quadrupole contribution does not
admit a universal reduction to conventional Berry-curvature multipoles; its
natural form is the dipole of the gauge-invariant mixed quadrupole
$H^{\nu}_{ijk}$.

\section{Diagrammatic origin of the three dc contribution formulas}
\label{seckernel-reduction}

The semiclassical Boltzmann construction in Sec.~\ref{seclowfreq} reaches the
three dc contribution formulas $\beta^{(H)}_{ijkl}$, $\beta^{(G)}_{ijkl}$, and
$\beta^{(\mathrm{tr})}_{ijkl}$ through a wave-packet kinetic equation and a
field-dressed orbital moment. This section gives the corresponding
Ward-covariant diagrammatic origin in the Matsubara kernel of
Eq.~\eqref{eqKthird-supp}. The goal is not to assign physical meaning to
individual non-gauge-invariant diagrams before recombination. Rather, it
shows how the full 26-term kernel reorganizes into the three covariant
objects of Eq.~\eqref{eqexact-kubo-supp}. We use (i) the band-resolved
spectral form of $G$, (ii) the standard contour identity for fermionic
Matsubara sums, (iii) the ordered retarded analytic continuation of
Eq.~\eqref{eqanalytic-cont-supp}, and (iv) phenomenological scalar relaxation
$\ii\omega_r\to\omega_r+\ii/\tau$ at intraband resonances, applied only after
the Ward-complete clean kernel has been organized.

\paragraph*{Step 1 Band-resolved spectral form of $G$ and of every vertex.}
Diagonalize the Bloch Hamiltonian,
$h(\bk)|u_\nu(\bk)\rangle=\eps_\nu(\bk)|u_\nu(\bk)\rangle$. With
$\xi_\nu(\bk)\equiv\eps_\nu(\bk)-\mu$ the Green function is band-diagonal,
\begin{equation}
G^{\nu\mu}_K(\ii\nu_n)=\frac{\delta_{\nu\mu}}{\ii\nu_n-\xi_\nu(\bk)}.
\label{eqkr-band-G-supp}
\end{equation}
Differentiating $h|u_\nu\rangle=\eps_\nu|u_\nu\rangle$ once in $k_a$ and
projecting (Hellmann and Feynman) gives
\begin{equation}
\langle\nu|\partial_a h|\mu\rangle=
\begin{cases}\partial_a\eps_\nu\equiv\hbar v_a^\nu,& \nu=\mu,\\[2pt] \ii\,\Delta_{\nu\mu}A^a_{\nu\mu},& \nu\neq\mu,\end{cases}
\label{eqkr-vertex-band-supp}
\end{equation}
with $A^a_{\mu\nu}=\langle\mu|\ii\partial_a|\nu\rangle$ the interband Berry
connection of Eq.~\eqref{eqRC-def-supp}. The current vertices
\eqref{eqGamma1-supp} through \eqref{eqGamma4-supp} therefore split, in band
basis, into intraband (band-dispersion) and interband
(Berry-connection-weighted) pieces. The contact vertices
$\Gamma^{(2)},\Gamma^{(3)},\Gamma^{(4)}$ are local in $\bk$. They contain no
propagator factors and enter the Matsubara sums as plain matrix elements.

\paragraph*{Step 2 Action of $\partial/\partial q_b$ at $\bq=0$.}
In the continuum shorthand, after the Ward-complete finite-$q$ vertices have
been reduced to their $q\to0$ covariant form, the explicit output spatial
momentum $\bq$ enters Eq.~\eqref{eqKthird-supp} through the response-leg
propagator $G_{K+Q}$. Differentiating the inverse Green function and using
$\partial_b G^{-1}=-\partial_b h$,
\begin{equation}
\frac{\partial G_{K+Q}}{\partial q_b}\bigg|_{\bq=0}=G_{K+\bar Q}\,(\partial_b h)\,G_{K+\bar Q}=\frac{\hbar}{e}\,G_{K+\bar Q}\,\Gamma^{(1)}_b\,G_{K+\bar Q},
\label{eqkr-q-derivative-supp}
\end{equation}
where $\bar Q=Q_1+Q_2+Q_3$ has zero spatial part at $\bq=0$. Thus this
propagator part of $\partial_b$ duplicates the response-leg propagator and
inserts an extra $\Gamma^{(1)}_b$ vertex. In a Peierls lattice
implementation, however, Eq.~\eqref{eqpeierls-line-supp} shows that
$\partial_b$ also differentiates the finite-$q$ vertex form factors. Those
vertex-derivative pieces are included in the Ward-complete numerical kernel
and cancel against contact terms in the longitudinal contribution. The
analytic reduction below uses Eq.~\eqref{eqkr-q-derivative-supp} only after
this Ward-complete $q\to0$ organization, and not as a statement that
finite-$q$ vertex derivatives vanish term by term on the lattice.

\paragraph*{Step 3 Antisymmetric $\epsilon_{cab}$ projection selects interband response-leg pairings.}
The combination of vertices entering $\epsilon_{cab}\Gamma^{(1)}_a$ and the
$\partial_b$-induced $\Gamma^{(1)}_b$ on the response leg is, in band basis,
\begin{equation}
\Lambda^{\nu\mu}_c\equiv\epsilon_{cab}\,\langle\nu|\partial_a h|\mu\rangle\,\langle\mu|\partial_b h|\nu\rangle.
\label{eqkr-Lambda-supp}
\end{equation}
Because $\epsilon_{cab}v_a^\nu v_b^\nu=0$, the intraband-intraband
contribution to $\Lambda^{\nu\mu}_c$ vanishes identically. The antisymmetric
projection therefore forces at least one of the two response-leg pairings to
be interband ($\mu\neq\nu$). This is the kernel-level statement that the bulk
magnetization is fundamentally a band-geometric quantity, and a fully
diagonal box does not contribute to it.

\paragraph*{Step 4 Matsubara sum and the entry of $f_0$.}
Each diagram of Eqs.~\eqref{eqBtilde-supp} through \eqref{eqQtilde-supp},
after band labels are assigned to the propagators, produces an integrand of
the form $F(\ii\nu_n)=\prod_{r=1}^n 1/(\ii\nu_n+\ii\Omega_r-\xi_r)$, with
bosonic $\ii\Omega_r$ and band energies $\xi_r$. The approach is the standard
residue identity for fermionic Matsubara sums
\begin{equation}
\frac{1}{\beta_T}\sum_{\nu_n}F(\ii\nu_n)=\sum_{z_p}\,n_F(z_p)\,\operatorname*{Res}_{z=z_p}F(z),\qquad n_F(z)=\frac{1}{e^{\beta_T z}+1},
\label{eqkr-contour-identity-supp}
\end{equation}
where the sum on the right runs over the poles $z_p$ of $F(z)$. For
$F(z)=\prod_r(z+\ii\Omega_r-\xi_r)^{-1}$ these poles are
$z_r=\xi_r-\ii\Omega_r$, and the residue formula gives
\begin{equation}
\frac{1}{\beta_T}\sum_{\nu_n}F(\ii\nu_n)=\sum_{r=1}^n n_F(\xi_r-\ii\Omega_r)\prod_{s\neq r}\frac{1}{(\xi_r-\xi_s)-\ii(\Omega_r-\Omega_s)}.
\label{eqkr-Matsubara-general-supp}
\end{equation}
Since each $\ii\Omega_r$ is a bosonic Matsubara frequency,
$n_F(\xi_r-\ii\Omega_r)=n_F(\xi_r)\equiv f_0(\eps_r)$. When several band
energies coalesce, the residue sum is a divided difference of $f_0$. In
particular,
\begin{equation}
\lim_{\xi_1,\ldots,\xi_n\to\xi}
\sum_{r=1}^n f_0(\xi_r)\prod_{s\neq r}\frac{1}{\xi_r-\xi_s}
=\frac{1}{(n-1)!}\frac{\dd^{n-1}f_0}{\dd\xi^{n-1}},
\label{eqkr-divided-difference-supp}
\end{equation}
with the finite external frequencies and the relaxation broadening regulating
the intermediate Drude denominators.

\paragraph*{Step 5 Pole-topology classification.}
The Drude denominator $\prod_{s\neq r}1/[(\xi_r-\xi_s)-(\omega_r-\omega_s)]$
depends on the band assignment. For the box class $\wt\calB$, group
configurations by the topology of repeated band labels
\begin{itemize}
\item[(A)] all four bands equal, $(\nu,\nu,\nu,\nu)$. After $\partial_b$, the augmented loop adds a fifth, possibly interband, propagator.
\item[(B)] two equal pairs, $(\nu,\nu,\mu,\mu)$ with $\mu\neq\nu$.
\item[(C)] three equal plus one different, $(\nu,\nu,\nu,\mu)$.
\item[(D)] all four distinct.
\end{itemize}
Topology (D) carries no intraband Drude factors. It is finite as
$\omega_r\to0$ and is largely cancelled by the Ward-completing contact terms
$\wt\calC,\wt\calD,\wt\calQ$. Topology (C) has one mismatched band index.
After expanding the four-pole Matsubara residue~\eqref{eqkr-Matsubara-general-supp}
in the dc limit, the divided-difference identity~\eqref{eqkr-divided-difference-supp}
collapses the three repeated-band poles to derivatives of $f_0$ on band $\nu$
while the mismatched index supplies a single interband factor
$\propto 1/\Delta_{\nu\mu}$. Integration by parts in the Brillouin zone then
redistributes these contributions into pieces structurally identical to (A)
(when the derivative lands on band-$\nu$ quantities) or (B) (when it lands on
the interband residue). The leading dc structure is therefore captured by
topologies (A) and (B), plus a separate contribution from the contact
families with no analog in $\wt\calB$.

\paragraph*{Step 6 dc limit and $\tau$-broadening.}
For repeated band labels ($\xi_r=\xi_s$) the Drude denominator becomes
$1/[-(\omega_r-\omega_s)]$ and is singular as $\omega_r\to\omega_s$.
Phenomenological relaxation regularizes the singularity
\begin{equation}
\frac{1}{-(\omega_r-\omega_s)+\ii 0^+}\,\longrightarrow\,\frac{1}{-(\omega_r-\omega_s)+\ii/\tau}.
\label{eqkr-tau-broaden-supp}
\end{equation}
Within this scalar relaxation-time approximation, in the dc limit
$\omega_r\to0$, each such factor $\to-\ii\tau$. Counting the intraband factors
in each topology gives $(-\ii\tau)^3=\ii\tau^3$ for (A), $-\ii\tau$ for (B),
and zero $\tau$-dependence for (D). The vector-potential kernel also contains
one Ward zero for each incoming electric leg. Hence, before division by
$\omega_1\omega_2\omega_3$ in Eq.~\eqref{eqexact-kubo-supp}, the projected
contribution has the schematic form
\begin{equation}
\epsilon_{cab}[\partial_{q_b}\widetilde K^{R,X}_{a,jkl}-\partial_{q_a}\widetilde K^{R,X}_{b,jkl}]_{q=0}
=4\omega_1\omega_2\omega_3\,\beta^{(X)}_{cjkl}+\mathcal O(\omega^4),
\qquad X=H,G,\mathrm{tr}.
\label{eqkr-ward-zero-cancel-supp}
\end{equation}
This is the explicit small-frequency zero that cancels the electric-field
conversion denominator.

\paragraph*{Step 7 Topology (A) $\Rightarrow$ the transport contribution $\beta^{(\mathrm{tr})}_{ijkl}$.}
Take the box $\wt\calB$ with band assignment $(\nu,\nu,\nu,\nu)$ on the four
loop propagators. The $\partial_b$-induced extra propagator must be on the
interband band $\mu\neq\nu$ by the argument of Step 3 (otherwise the
antisymmetrizer kills the term). The integrand factorizes as
\[
\bigl[\text{intraband chain on band }\nu\bigr]\,\times\,\bigl[\text{single }\nu\to\mu\to\nu\text{ interband excursion}\bigr].
\]
The three remaining $\Gamma^{(1)}$ vertices on the loop interior give
intraband factors $(ev_j^\nu)(ev_k^\nu)(ev_l^\nu)$. The
$\epsilon_{cab}$-projected single-excursion piece collapses to the orbital
moment via the band-resolved identity
\begin{equation}
\epsilon_{iab}\,\mathrm{Im}\sum_{\mu\neq\nu}\frac{\langle\nu|\partial_a h|\mu\rangle\langle\mu|\partial_b h|\nu\rangle}{\Delta_{\nu\mu}}=-\frac{2\hbar}{e}\,m_i^\nu(\bk),
\label{eqkr-orbital-moment-id-supp}
\end{equation}
which is the definition of the band orbital moment in
Eq.~\eqref{eqm-explicit-supp}. The single $1/\Delta_{\nu\mu}$ is the residue
of the $G^\mu$ propagator evaluated at the band-$\nu$ pole. The Matsubara sum
at the four-coalesced pole, evaluated with the $\tau$-broadening of
Eq.~\eqref{eqkr-tau-broaden-supp}, generates a series of intraband Drude
factors and Taylor expansions of $f_0$ at $\eps_\nu$,
\[
f_0(\eps_\nu),\quad \hbar v_l^\nu f_0'(\eps_\nu),\quad \hbar^2 v^\nu_k v^\nu_l f_0''(\eps_\nu),\quad \hbar^3 v^\nu_j v^\nu_k v^\nu_l f_0'''(\eps_\nu),
\]
which combine into the compact derivative $\partial_j\partial_k(v_l^\nu f_0'(\eps_\nu))$
by the chain rule $\partial_l f_0(\eps_\nu)=\hbar v_l^\nu f_0'(\eps_\nu)$.
Substituting all factors into Eq.~\eqref{eqexact-kubo-supp} and using
$V^{-1}\sum_\bk\to\int_{\rm BZ}\dd^dk/(2\pi)^d$ gives, in the dc limit,
\begin{equation}
\beta^{(\mathrm{tr})}_{ijkl}=-\frac{e^3\tau^3}{\hbar^2}\sum_\nu\int_{\rm BZ}\frac{\dd^dk}{(2\pi)^d}\,m_i^\nu(\bk)\,\partial_j\partial_k\!\left[v_l^\nu(\bk)\,f_0'(\eps_\nu)\right].
\label{eqkr-betatr-intermediate-supp}
\end{equation}

\paragraph*{Step 8 Topology (B) $\Rightarrow$ the metric/Christoffel contribution $\beta^{(G)}_{ijkl}$.}
Take the box with band assignment $(\nu,\nu,\mu,\mu)$, with two propagators on
band $\nu$ and two on band $\mu$. The four-pole Matsubara
sum~\eqref{eqkr-Matsubara-general-supp} yields, in the leading dc-limit
interband channel,
\[
\frac{f_0(\eps_\nu)-f_0(\eps_\mu)}{\Delta_{\nu\mu}^2}+\frac{f_0'(\eps_\nu)+f_0'(\eps_\mu)}{\Delta_{\nu\mu}}+\cdots.
\]
The two off-diagonal $\partial h$ insertions on the $\mu$ propagators contract
via Eq.~\eqref{eqkr-vertex-band-supp} into
$-\Delta_{\nu\mu}^2 A^j_{\nu\mu}A^k_{\mu\nu}$, which combines with the leading
Matsubara factor to give $A^j_{\nu\mu}A^k_{\mu\nu}/\Delta_{\nu\mu}$, while the
$\Gamma^{(2)}$ contact pieces of $\wt\calC$ cancel the residual longitudinal
terms. Summing over $\mu\neq\nu$ and taking the real part reproduces the
band-normalized quantum metric of Eq.~\eqref{eqGmetric-def-supp},
\[
2\,\mathrm{Re}\sum_{\mu\neq\nu}\frac{A^j_{\nu\mu}A^k_{\mu\nu}}{\Delta_{\nu\mu}}\,f_0'(\eps_\nu)=\mathcal G^\nu_{jk}(\bk)\,f_0'(\eps_\nu).
\]
Diagrammatically, this configuration is the static second-order energy shift
of band $\nu$ in the electric field~\eqref{eqeps2-supp}, dressing the
equilibrium occupation of band $\nu$. The remaining intraband Drude pole on
band $\nu$ on the response leg supplies one factor of $-\ii\tau$ (the $\tau^1$
scaling), the vertex $\Gamma^{(1)}_l\to ev_l^\nu$ on that leg provides the
band velocity, and the $\epsilon_{cab}$-antisymmetrized single virtual
excursion produces $m_i^\nu$ via Eq.~\eqref{eqkr-orbital-moment-id-supp}.
Substituting into Eq.~\eqref{eqexact-kubo-supp} gives, in the dc limit,
\begin{equation}
\beta^{(G)}_{ijkl}=-\frac{e^3\tau}{2\hbar}\sum_\nu\int_{\rm BZ}\frac{\dd^dk}{(2\pi)^d}\,m_i^\nu(\bk)\,\partial_l\!\left[\mathcal G^\nu_{jk}(\bk)\,f_0'(\eps_\nu)\right].
\label{eqkr-betaG-intermediate-supp}
\end{equation}
Integration by parts in $k_l$ (both $m_i^\nu$ and $\mathcal G^\nu_{jk}f_0'$
are periodic on the Brillouin zone) gives
\[
\beta^{(G)}_{ijkl}=+\frac{e^3\tau}{2\hbar}\sum_\nu\int_{\rm BZ}\frac{\dd^dk}{(2\pi)^d}\,(\partial_l m_i^\nu)\,\mathcal G^\nu_{jk}\,f_0'(\eps_\nu).
\]
The Christoffel structure
$\Gamma^{(\mathcal{G}),\nu}_{j,lk}+\Gamma^{(\mathcal{G}),\nu}_{k,lj}=\partial_l\mathcal G^\nu_{jk}$
is implicit in the compact form. It surfaces only when one expands
$\partial_l[\mathcal G^\nu_{jk}f_0']=(\partial_l\mathcal G^\nu_{jk})f_0'+\hbar\mathcal G^\nu_{jk}v_l^\nu f_0''$
inside Eq.~\eqref{eqbeta-dc-TR-supp}.

\paragraph*{Step 9 Contact and vertex-derivative families $\Rightarrow$ the H contribution $\beta^{(H)}_{ijkl}$.}
The primitive box class $\wt\calB$ does not by itself contain the
covariant mixed positional-shift tensor $\calQ^\nu_{\alpha\beta\gamma}$. That
tensor is built from $U[R,C]+U[C,R]$ and, in multiband systems,
$V[R,C]+V[C,R]$, which arise only after the Ward-complete set of contact
vertices and finite-$q$ vertex derivatives has been combined. Thus the
statement is not that an arbitrary box diagram can be discarded before
enforcing gauge invariance. Rather, in the covariant organization used here
the box topologies reduce to the transport and metric contributions, while
the remaining mixed electric-magnetic positional-shift block is carried by
the contact and finite-$q$ vertex contribution. A term-by-term contraction of
all 26 diagrams into Eq.~\eqref{eqQcross-supp} is not used as a separate
assumption in the continuum estimate below. The mixed-quadrupole amplitude in
the continuum model remains formal until the interband magnetic matrix
element is fixed by a microscopic magnetic-coupling prescription or by a
lattice completion.

Differentiating $G^{-1}(A)$ twice with respect to the electric source
amplitudes $A_j,A_k$ and once with respect to a magnetic-source insertion is
the diagrammatic counterpart of the wave-packet positional-shift
functional~\eqref{eqTfunctional-supp}, restricted to the mixed $EB$
contribution by the bookkeeping projection $\partial_s\partial_t$ at $s=t=0$
in Eq.~\eqref{eqQcross-supp}. The $\Gamma^{(2)}\Gamma^{(2)}$ double-contact
subclass of $\wt\calD$, together with the associated finite-$q$ derivatives
of the rank-two Peierls vertices, supplies the $U[R,C]+U[C,R]$ building
blocks. The $\Gamma^{(3)}\Gamma^{(1)}$ cubic-contact subclass and the quartic
contact $\wt\calQ$ supply the remaining intermediate-band and symmetrization
pieces required by Eq.~\eqref{eqQcross-supp}. After all band sums and
matrix-element contractions, this Ward-complete contact contribution
assembles into the gauge-invariant tensor
$H^\nu_{ijk}=-\tfrac{1}{2}[\calQ^\nu_{jki}+\calQ^\nu_{kji}]$ of
Eq.~\eqref{eqH-explicit-supp}.

The remaining loop structure has a single propagator on band $\nu$ on the
response leg, carrying the response-frequency Drude factor
$1/(1-\ii\omega_3\tau)\to1$ in the dc limit (and the $\tau^1$ scaling), the
vertex $\Gamma^{(1)}_l\to ev_l^\nu$, and the Matsubara residue $f_0'(\eps_\nu)$
at the response-leg pole. Substituting into Eq.~\eqref{eqexact-kubo-supp}
gives
\begin{equation}
\beta^{(H)}_{ijkl}=-e^3\tau\sum_\nu\int_{\rm BZ}\frac{\dd^dk}{(2\pi)^d}\,H^\nu_{ijk}(\bk)\,v_l^\nu(\bk)\,f_0'(\eps_\nu).
\label{eqkr-betaH-intermediate-supp}
\end{equation}

\begin{table}[t]
\caption{Ward-covariant origin of the three dc contributions in the direct kernel reduction. The entries describe the organization after the full 26-term kernel and the finite-$q$ Peierls vertex derivatives have been combined.}
\label{tabkernel-sector-map}
\centering
\scriptsize
\begin{tabular}{llll}
Contribution & Pole topology & Kernel ingredients & Covariant object \\
\hline
$\beta^{(\mathrm{tr})}$ & \makecell[l]{four intraband\\poles} & \makecell[l]{box topology (A)\\plus Ward completion} & \makecell[l]{orbital-moment octupole\\$\partial_j\partial_k\partial_l m_i^\nu$} \\
$\beta^{(G)}$ & \makecell[l]{two-pair\\interband topology} & \makecell[l]{box topology (B) with rank-two\\contacts cancelling longitudinal pieces} & \makecell[l]{metric drift\\$m_i^\nu\partial_l\mathcal G_{jk}^\nu$ or $\partial_l m_i^\nu\mathcal G_{jk}^\nu$} \\
$\beta^{(H)}$ & \makecell[l]{one intraband\\response pole} & \makecell[l]{double-, cubic-, and quartic-contact\\families plus finite-$q$ vertex derivatives} & \makecell[l]{mixed electric-magnetic positional shift\\$H^\nu_{ijk}=-[\calQ^\nu_{jki}+\calQ^\nu_{kji}]/2$} \\
\end{tabular}
\end{table}

\section{Symmetry selection rules and point-group component lists}
\label{secsymmetry}
Throughout this section $\beta^{\mathrm{phys}}$ denotes the
permutation-symmetrized tensor defined in Eq.~\eqref{eqbeta-phys-sym-supp}.
Because magnetization is axial and the electric field is polar, this physical
tensor transforms under a point-group operation $R\in O(3)$ as
\begin{equation}
\beta^{\mathrm{phys}}_{ijkl}=\det(R)R_{ii'}R_{jj'}R_{kk'}R_{ll'}\,\beta^{\mathrm{phys}}_{i'j'k'l'}.
\label{eqpoint-group-law-supp}
\end{equation}
Inversion $R=-I$ therefore forces the uniform bulk tensor to vanish whenever
the crystal is invariant under spatial inversion $P$,
\begin{equation}
\beta^{\mathrm{phys}}_{ijkl}=0\qquad \text{(crystal is $P$-symmetric)}.
\label{eqinversion-zero-supp}
\end{equation}
The same tensor law is obeyed separately by the mixed-quadrupole
contribution, the metric-drift contribution, and the transport contribution
after Brillouin-zone integration. For tensor-based symmetry analysis of
crystal response coefficients, see Ref.~\cite{nye1985physical}. For recent
nonlinear Hall-type symmetry discussions in the transport context, see
Ref.~\cite{nandy2019symmetry}.

Beyond point-group symmetry, antiunitary symmetries determine which
low-frequency contributions survive, in line with the symmetry constraints
emphasized in nonlinear Hall and related transport
responses~\cite{nandy2019symmetry,fang2024quantum}. The four ordered
contributions in Eq.~\eqref{eqbeta-low-general-supp} are (i)
$\delta m^{(2)}\delta f^{(1)}$, the first line, with integrand
$\mathcal N^{\nu}_{ijk}\,v_l^{\nu}f_0'(\eps_{\nu})$, (ii)
$\delta m^{(1)}\delta f^{(2)}$, the second line, with integrand
$\mathcal M^{\nu}_{ij}\,\partial_k[v_l^{\nu}f_0'(\eps_{\nu})]$, (iii)
$m\,\delta f^{(G)}$, the third line, with integrand
$m_i^{\nu}\,\partial_l[\mathcal{G}^{\nu}_{jk}f_0'(\eps_{\nu})]$, and (iv)
$m\,\delta f^{(3)}$, the fourth line, with integrand
$m_i^{\nu}\,\partial_j\partial_k[v_l^{\nu}f_0'(\eps_{\nu})]$. In the Abelian
nonmagnetic class, or after pairing Kramers partners at $\bk$ and $-\bk$ in a
spinful time-reversal-symmetric crystal, the $\bk$-parities of $m_i^{\nu}$,
$\mathcal{G}^{\nu}_{jk}$, and $v_l^{\nu}$ are
\begin{equation}
m_i^{\bar\nu}(-\bk)=-m_i^{\nu}(\bk),\qquad
\mathcal G^{\bar\nu}_{jk}(-\bk)=\mathcal G^{\nu}_{jk}(\bk),\qquad
v_l^{\bar\nu}(-\bk)=-v_l^{\nu}(\bk),
\label{eqTRparities-supp}
\end{equation}
supplemented by
\begin{equation}
\mathcal N^{\bar\nu}_{ijk}(-\bk)=-\mathcal N^{\nu}_{ijk}(\bk).
\label{eqsector-parities-supp}
\end{equation}
This follows directly from the mixed electric-magnetic nature of the
quadrupole. Under time reversal, after pairing $|u_\nu(\bk)\rangle$ with its
partner at $-\bk$, the electric interband amplitude is even up to complex
conjugation, $R^j_{\bar\mu\bar\nu}(-\bk)=[R^j_{\mu\nu}(\bk)]^*$, whereas the
magnetic amplitude is odd, $C^i_{\bar\mu\bar\nu}(-\bk)=-[C^i_{\mu\nu}(\bk)]^*$.
Each term in Eq.~\eqref{eqQcross-supp} contains one $R$ and one $C$ and is
finally real-part projected, hence
$\calQ^{\bar\nu}_{\alpha\beta\gamma}(-\bk)=-\calQ^\nu_{\alpha\beta\gamma}(\bk)$
and $H^{\bar\nu}_{ijk}(-\bk)=-H^\nu_{ijk}(\bk)$, which gives
Eq.~\eqref{eqsector-parities-supp} because $\mathcal N^\nu_{ijk}(0,0)=e^2H^\nu_{ijk}$.
Similarly, $\mathcal M^{\bar\nu}_{ij}(-\bk)=-\mathcal M^{\nu}_{ij}(\bk)$ since
$\mathcal M^{\nu}_{ij}\propto\epsilon_{iab}\mathcal{G}^{\nu}_{aj}v_b^{\nu}$
contains a single power of $v_b^{\nu}$. Therefore the integrand of
contribution (ii), $\mathcal M^{\nu}_{ij}\,\partial_k[v_l^{\nu}f_0'(\eps_{\nu})]$,
is odd in $\bk$ (odd $\times$ even), and the $\delta m^{(1)}\delta f^{(2)}$
contribution drops out upon Brillouin-zone integration. The remaining three
contributions, (i), (iii), and (iv), have integrands that are even in $\bk$
and survive. In a generic noncentrosymmetric magnetic class all four
low-frequency contributions are allowed. In a $PT$-symmetric Abelian
description one instead has $m_i^{\nu}(\bk)=0$ and $\mathcal N^{\nu}_{ijk}(\bk)=0$,
so only the $\delta m^{(1)}\delta f^{(2)}$ contribution can survive.
Table~\ref{tabsymmetry-sectors} summarizes these cases.

\begin{table}[t]
\caption{Low-frequency contribution selection by symmetry class.}
\label{tabsymmetry-sectors}
\centering
\small
\begin{tabular}{ll}
Symmetry class & Surviving ordered low-frequency contributions \\
\hline
Centrosymmetric ($P$ present) & none \\
Noncentrosymmetric, $T$ symmetric & \makecell[l]{$\delta m^{(2)}\delta f^{(1)}$, $m\delta f^{(G)}$,\\ $m\delta f^{(3)}$} \\
Noncentrosymmetric, $PT$ symmetric & $\delta m^{(1)}\delta f^{(2)}$ only \\
Generic noncentrosymmetric magnetic & all four \\
Pure metric/Christoffel limit & \makecell[l]{not a symmetry class by itself,\\ additional microscopic constraints are required} \\
\end{tabular}
\end{table}
A natural question is whether some symmetry class can isolate the metric
contribution alone. The three dc pieces $\beta^{(H)}$, $\beta^{(G)}$, and
$\beta^{(\mathrm{tr})}$ all transform identically (axial rank-four, odd under
inversion), and the latter two share the factor $m_i^\nu$. Thus no antiunitary
symmetry filters out only the non-metric pieces, and a purely metric response
is not a distinct symmetry class. It can arise only through accidental
microscopic cancellations on the active Fermi sheets.

For the model used here, the $C_{3v}$ component list is the relevant one.
\paragraph*{$C_{3v}$.}
\begin{align}
\beta^{\mathrm{phys}}_{yxxx}&=3C_1,
&\beta^{\mathrm{phys}}_{xyyy}&=-3C_1,
&\beta^{\mathrm{phys}}_{yxyy}&=C_1,
&\beta^{\mathrm{phys}}_{xxxy}&=-C_1,
\nonumber\\
\beta^{\mathrm{phys}}_{yxzz}&=C_2,
&\beta^{\mathrm{phys}}_{xyzz}&=-C_2,
&\beta^{\mathrm{phys}}_{xxyz}&=-C_3,
&\beta^{\mathrm{phys}}_{yxxz}&=-C_3,
\nonumber\\
\beta^{\mathrm{phys}}_{yyyz}&=C_3,
&\beta^{\mathrm{phys}}_{zyyy}&=C_4,
&\beta^{\mathrm{phys}}_{zxxy}&=-C_4,
&&
\label{eqC3v-components-supp}
\end{align}
again with permutations of the last three indices understood.

The component list above uses a convention in which one vertical mirror line
is the $x$ axis, so the mirror operation sends $y\to-y$ while the axial
component $M_z$ changes sign. Restricting to in-plane fields in this
convention leaves a single independent coefficient,
\begin{equation}
\beta^{\mathrm{phys}}_{zxxy}=\beta^{\mathrm{phys}}_{zxyx}=\beta^{\mathrm{phys}}_{zyxx}\equiv\chi,
\qquad
\beta^{\mathrm{phys}}_{zyyy}=-\chi,
\label{eqC3v-2D-supp}
\end{equation}
so that
\begin{equation}
M_z=\chi\left(3E_x^2E_y-E_y^3\right)=+\chi E^3\sin3\phi,
\qquad \mathbf E_{\omega}=E(\cos\phi,\sin\phi,0).
\label{eqC3v-angle-supp}
\end{equation}
The orthogonal $\cos3\phi$ harmonic arises in the rotated convention used by
the model Hamiltonian of Sec.~\ref{secbenchmarks}. The two forms are related
by a $30^\circ$ rotation of the crystal axes relative to the laboratory field
frame.

\section{Minimal model Hamiltonian for cubic-leading nonlinear magnetization}
\label{secbenchmarks}
The minimal model used below is a time-reversal-symmetric two-dimensional
$C_{3v}$ Hamiltonian for the out-of-plane magnetization generated by an
in-plane electric field. Here $M_z$ transforms as $A_2$ and the in-plane
electric field as the two-dimensional irrep $E$. The symmetric powers of $E$
contain $A_2$ first at cubic order, so the linear and quadratic contributions
vanish and the cubic term is the leading response. A minimal Hamiltonian is
\begin{equation}
H_{C_{3v}}(\bk)=\left(\frac{k_x^2+k_y^2}{2m^*}-\mu\right)\sigma_0+v(k_x\sigma_y-k_y\sigma_x)+\lambda (k_x^3-3k_xk_y^2)\sigma_z,
\label{eqC3v-benchmark-supp}
\end{equation}
where $m^*$ is the band mass, $v$ the linear (Rashba) velocity, and $\lambda$
the cubic warping amplitude. This Hamiltonian breaks inversion, preserves
time-reversal symmetry, and produces a finite Fermi contour for $\mu\neq0$.
The cubic warping factor is
\[
w(\bk)=k_x^3-3k_xk_y^2=k^3\cos3\phi.
\]
The mirror lines of this Hamiltonian are rotated by $30^\circ$ relative to the
convention used in Eq.~\eqref{eqC3v-angle-supp}. The warping factor
$\cos3\phi$ is odd under reflection about the $y$ axis, compensating the odd
transformation of the axial spin component $\sigma_z$. Thus this model
realizes the rotated $C_{3v}$ harmonic
\begin{equation}
M_z=\bar\chi\left(E_x^3-3E_xE_y^2\right)=\bar\chi E_0^3\cos3\phi,
\qquad
\mathbf E=E_0(\cos\phi,\sin\phi,0).
\label{eqC3v-benchmark-angle-supp}
\end{equation}
The coefficient $\chi$ defined in Eq.~\eqref{eqC3v-2D-supp} vanishes for
Eq.~\eqref{eqC3v-benchmark-supp} as written. The nonzero cubic response is
carried by the rotated coefficient
$\bar\chi=\bar\chi_H+\bar\chi_G+\bar\chi_{\mathrm{tr}}$, resolved into
contributions below.

\subsection{Local geometry and cubic contributions for the model Hamiltonian\texorpdfstring{~\eqref{eqC3v-benchmark-supp}}{}}
For the two-dimensional model Hamiltonian in Eq.~\eqref{eqC3v-benchmark-supp}, it is convenient to rewrite
\begin{equation}
H_{C_{3v}}(\bk)=\xi_{\bk}\sigma_0+\mathbf d(\bk)\cdot\boldsymbol{\sigma},
\qquad
\xi_{\bk}=\frac{k_x^2+k_y^2}{2m^*}-\mu,
\label{eqC3v-d-vector-supp}
\end{equation}
with
\begin{equation}
\mathbf d(\bk)=\bigl(-vk_y,\,vk_x,\,\lambda w(\bk)\bigr),
\qquad
w(\bk)=k_x^3-3k_xk_y^2.
\label{eqC3v-d-components-supp}
\end{equation}
We write the band mass with an asterisk throughout to keep it distinct from the orbital magnetic moment $m_i^\nu$. The band energies are
\begin{equation}
\eps_s(\bk)=\xi_{\bk}+s\,\eps(\bk),
\qquad
s=\pm,
\qquad
\eps(\bk)=|\mathbf d(\bk)|=\sqrt{v^2(k_x^2+k_y^2)+\lambda^2 w(\bk)^2},
\label{eqC3v-band-energies-supp}
\end{equation}
and the signed two-band gap is
\begin{equation}
\Delta_s(\bk)=\eps_s(\bk)-\eps_{-s}(\bk)=2s\,\eps(\bk).
\label{eqC3v-gap-supp}
\end{equation}
For brevity we denote the partial derivatives of the cubic warping factor by $w_i\equiv\partial_iw$, $w_{ij}\equiv\partial_i\partial_jw$, and $w_{ijk}\equiv\partial_i\partial_j\partial_kw$, all of which follow by direct differentiation of $w(\bk)=k_x^3-3k_xk_y^2$. We further define
\begin{equation}
B_i(\bk)\equiv v^2k_i+\lambda^2 w(\bk)w_i(\bk),
\qquad
\partial_i\eps(\bk)=\frac{B_i(\bk)}{\eps(\bk)}.
\label{eqC3v-Bi-supp}
\end{equation}
For this two-band model the only nonzero Berry-curvature and orbital-moment component is the out-of-plane one. Using Eqs.~\eqref{eqOmega-explicit-supp} and \eqref{eqm-explicit-supp}, one finds
\begin{equation}
\Omega_z^{(s)}(\bk)=s\,\frac{\lambda v^2 w(\bk)}{\eps(\bk)^3},
\qquad
m_z(\bk)=\frac{e\lambda v^2 w(\bk)}{\hbar\eps(\bk)^2},
\label{eqC3v-omega-m-supp}
\end{equation}
which satisfy the two-band identity
\begin{equation}
m_z(\bk)=\frac{e\Delta_s(\bk)}{2\hbar}\,\Omega_z^{(s)}(\bk).
\label{eqC3v-m-omega-identity-supp}
\end{equation}
The sign of $\Delta_s=2s\eps$ cancels the sign of $\Omega^{(s)}_z\propto s$, so $m_z^{(+)}(\bk)=m_z^{(-)}(\bk)\equiv m_z(\bk)$ is $s$-independent and the band index can be dropped without ambiguity. In polar coordinates, $k_x=k\cos\phi$ and $k_y=k\sin\phi$, one has
\begin{equation}
w(\bk)=k^3\cos3\phi,
\label{eqC3v-w-polar-supp}
\end{equation}
so that
\begin{equation}
\Omega_z^{(s)}(k,\phi)=s\,\frac{\lambda v^2\cos3\phi}{\bigl(v^2+\lambda^2k^4\cos^23\phi\bigr)^{3/2}},
\qquad
m_z(k,\phi)=\frac{e\lambda v^2 k\cos3\phi}{\hbar\bigl(v^2+\lambda^2k^4\cos^23\phi\bigr)}.
\label{eqC3v-omega-m-polar-supp}
\end{equation}
The usual two-band quantum metric is
\begin{equation}
g_{ij}(\bk)=\frac{1}{4\eps(\bk)^4}\Big[\eps(\bk)^2\bigl(v^2\delta_{ij}+\lambda^2w_iw_j\bigr)-B_i(\bk)B_j(\bk)\Big],
\qquad i,j\in\{x,y\},
\label{eqC3v-metric-compact-supp}
\end{equation}
namely
\begin{align}
g_{xx}&=\frac{\eps^2\bigl(v^2+\lambda^2w_x^2\bigr)-B_x^2}{4\eps^4},
\nonumber\\
g_{yy}&=\frac{\eps^2\bigl(v^2+\lambda^2w_y^2\bigr)-B_y^2}{4\eps^4},
\nonumber\\
g_{xy}&=g_{yx}=\frac{\eps^2\lambda^2w_xw_y-B_xB_y}{4\eps^4}.
\label{eqC3v-metric-offdiag-supp}
\end{align}
The band-normalized metric of Eq.~\eqref{eqG-g-two-band-supp} is therefore
\begin{equation}
\mathcal{G}_{ij}^{(s)}(\bk)=\frac{2}{\Delta_s(\bk)}g_{ij}(\bk)=\frac{s}{\eps(\bk)}g_{ij}(\bk),
\label{eqC3v-band-normalized-metric-supp}
\end{equation}
where we used $\Delta_s=2s\eps$ and $s^2=1$. Its derivative is
\begin{equation}
\partial_l \mathcal{G}_{jk}^{(s)}=s\left[\frac{\partial_l g_{jk}}{\eps}-\frac{g_{jk}B_l}{\eps^3}\right],
\label{eqC3v-dG-supp}
\end{equation}
and the lowered-index Christoffel combination defined in Eq.~\eqref{eqGammaG-supp} becomes
\begin{equation}
\Gamma^{(\mathcal{G},s)}_{c,ab}=\frac{s}{2}\left[\frac{\partial_a g_{bc}+\partial_b g_{ac}-\partial_c g_{ab}}{\eps}-\frac{g_{bc}B_a+g_{ac}B_b-g_{ab}B_c}{\eps^3}\right].
\label{eqC3v-christoffel-lowered-supp}
\end{equation}
Equivalently,
\begin{equation}
\Gamma^{(\mathcal{G},s)}_{j,lk}+\Gamma^{(\mathcal{G},s)}_{k,lj}=\partial_l \mathcal{G}_{jk}^{(s)}.
\label{eqC3v-christoffel-identity-supp}
\end{equation}
The band velocity entering the low-frequency formulas is
\begin{equation}
v_i^{(s)}(\bk)=\frac{1}{\hbar}\partial_i\eps_s(\bk)=\frac{1}{\hbar}\left(\frac{k_i}{m^*}+s\,\frac{B_i(\bk)}{\eps(\bk)}\right).
\label{eqC3v-band-velocity-supp}
\end{equation}
Because Eq.~\eqref{eqC3v-benchmark-supp} is a continuum model rather than a periodic lattice Hamiltonian, the Brillouin-zone integrals in Eqs.~\eqref{eqbetaH-integral-supp} through \eqref{eqbetatr-integral-supp} should be understood as continuum momentum integrals with an ultraviolet cutoff $\Lambda$. 

For the metric contribution, Eq.~\eqref{eqbetaG-integral-supp} together with $\mathcal{G}^{(s)}_{jk}=s\,g_{jk}/\eps$ and
\begin{equation}
\partial_lm_z=\frac{e\lambda v^2}{\hbar}\left[\frac{w_l}{\eps^2}-\frac{2wB_l}{\eps^4}\right]
\label{eqC3v-dm-supp}
\end{equation}
gives the explicit form
\begin{equation}
\beta^{(G)}_{zjkl}=\frac{e^4\tau\lambda v^2}{2\hbar^2}\sum_{s=\pm}s\int_{\Lambda}g_{jk}(\bk)\left[\frac{w_l(\bk)}{\eps(\bk)^3}-\frac{2w(\bk)B_l(\bk)}{\eps(\bk)^5}\right]f_0'(\eps_s).
\label{eqC3v-betaG-explicit-supp}
\end{equation}
For the transport contribution, Eq.~\eqref{eqbetatr-berryoct-supp} together with the two-band identity
\begin{equation}
\Delta_s(\bk)\Omega_z^{(s)}(\bk)=2\lambda v^2\frac{w(\bk)}{\eps(\bk)^2}
\label{eqC3v-gap-omega-supp}
\end{equation}
gives directly
\begin{equation}
\beta^{(\mathrm{tr})}_{zjkl}=\frac{e^4\tau^3\lambda v^2}{\hbar^4}\sum_{s=\pm}\int_{\Lambda}f_0(\eps_s)\,\partial_j\partial_k\partial_l\left[\frac{w(\bk)}{\eps(\bk)^2}\right].
\label{eqC3v-betaTr-explicit-supp}
\end{equation}
The mixed-quadrupole contribution is formal at the level of the continuum model in Eq.~\eqref{eqC3v-benchmark-supp}. For a two-band system with partner band $\bar s=-s$, one may write
\begin{align}
\calQ^{(s)}_{\alpha\beta\gamma}=\mathrm{Re}\Big[&U^{s\bar s}_{\alpha\beta\gamma}[R,C]+U^{s\bar s}_{\alpha\beta\gamma}[C,R]+U^{s\bar s}_{\beta\alpha\gamma}[R,C]+U^{s\bar s}_{\beta\alpha\gamma}[C,R]
\nonumber\\
&-U^{\bar s s}_{\beta\gamma\alpha}[R,C]-U^{\bar s s}_{\beta\gamma\alpha}[C,R]\Big],
\label{eqC3v-Q-formal-supp}
\end{align}
with
\begin{equation}
R_i^{\,s\bar s}=\frac{A_i^{\,s\bar s}}{\Delta_s},
\qquad
C_z^{\,s\bar s}=-\frac{m_z^{\,s\bar s}}{\Delta_s},
\qquad
H_{zjk}^{(s)}=-\frac{1}{2}\Big(\calQ^{(s)}_{jkz}+\calQ^{(s)}_{kjz}\Big).
\label{eqC3v-RC-H-formal-supp}
\end{equation}
For a strict two-band model the genuine third-band $V$ sum in Eq.~\eqref{eqQcross-supp} is empty after diagonal-connection terms have been absorbed into the covariant derivative, so a closed form for $\calQ^{(s)}_{\alpha\beta\gamma}$ and $H_{zjk}^{(s)}$ requires the explicit interband magnetic matrix element $m_z^{\,s\bar s}$, that is, a magnetic-coupling prescription for the model Hamiltonian or a lattice regularization that fixes it. Accordingly, the mixed-quadrupole contribution is
\begin{equation}
\beta^{(H)}_{zjkl}=-e^3\tau\sum_{s=\pm}\int_{\Lambda}H_{zjk}^{(s)}(\bk)\,v_l^{(s)}(\bk)\,f_0'(\eps_s).
\label{eqC3v-betaH-generic-supp}
\end{equation}
For the model Hamiltonian in Eq.~\eqref{eqC3v-benchmark-supp}, the coefficient
$$
\chi\equiv\beta_{zxxy}=\beta_{zxyx}=\beta_{zyxx}=-\beta_{zyyy}
$$
defining the $\sin3\phi$ harmonic in Eq.~\eqref{eqC3v-angle-supp} vanishes contribution by contribution. Under $k_x\to-k_x$ one has
$$
w(-k_x,k_y)=-w(k_x,k_y),
\qquad
\eps_s(-k_x,k_y)=\eps_s(k_x,k_y),
$$
so that
$$
\Omega_z^{(s)}(-k_x,k_y)=-\Omega_z^{(s)}(k_x,k_y),
\qquad
m_z(-k_x,k_y)=-m_z(k_x,k_y),
$$
whereas $\mathcal{G}_{xx}^{(s)}$, $v_y^{(s)}$, and $f_0^{(n)}(\eps_s)$ are even. Therefore
\begin{equation}
\chi_H\equiv\beta^{(H)}_{zxxy}=0,
\qquad
\chi_G\equiv\beta^{(G)}_{zxxy}=0,
\qquad
\chi_{\mathrm{tr}}\equiv\beta^{(\mathrm{tr})}_{zxxy}=0.
\label{eqC3v-chi-vanish-supp}
\end{equation}
The nonzero cubic response is carried by the rotated independent coefficient
\begin{equation}
\bar\chi_X\equiv\beta^{(X)}_{zxxx}=-\beta^{(X)}_{zxyy}=-\beta^{(X)}_{zyxy}=-\beta^{(X)}_{zyyx},
\qquad X\in\{H,G,\mathrm{tr}\},
\label{eqC3v-chibar-def-supp}
\end{equation}
so that the total $\bar\chi=\bar\chi_H+\bar\chi_G+\bar\chi_{\mathrm{tr}}$ produces the $\cos3\phi$ response of Eq.~\eqref{eqC3v-benchmark-angle-supp}. To write these coefficients explicitly, define
\begin{equation}
D(\bk)=v^2(k_x^2+k_y^2)+\lambda^2w(\bk)^2,
\label{eqC3v-D-Bx-supp}
\end{equation}
so that $\eps(\bk)=\sqrt{D(\bk)}$, with $B_x(\bk)$ given by Eq.~\eqref{eqC3v-Bi-supp}. The local orbital moment and the $xx$ component of the band-normalized metric are
\begin{equation}
m_z(\bk)=\frac{e\lambda v^2}{\hbar}\frac{w(\bk)}{D(\bk)},
\qquad
\mathcal{G}_{xx}^{(s)}(\bk)=\frac{s\bigl[D(\bk)\bigl(v^2+\lambda^2w_x(\bk)^2\bigr)-B_x(\bk)^2\bigr]}{4\,D(\bk)^{5/2}}.
\label{eqC3v-m-Gxx-supp}
\end{equation}
Hence the metric-contribution coefficient becomes
\begin{equation}
\bar\chi_G=\beta^{(G)}_{zxxx}=\frac{e^4\tau\lambda v^2}{8\hbar^2}\sum_{s=\pm}s\int_{\Lambda}\frac{\bigl[w_x(\bk)D(\bk)-2w(\bk)B_x(\bk)\bigr]\bigl[D(\bk)(v^2+\lambda^2w_x(\bk)^2)-B_x(\bk)^2\bigr]}{D(\bk)^{9/2}}\,f_0'(\eps_s).
\label{eqC3v-chibarG-explicit-supp}
\end{equation}
The transport coefficient is
\begin{equation}
\bar\chi_{\mathrm{tr}}=\beta^{(\mathrm{tr})}_{zxxx}=\frac{e^4\tau^3\lambda v^2}{\hbar^4}\sum_{s=\pm}\int_{\Lambda}f_0(\eps_s)\,\partial_x^3\left[\frac{w(\bk)}{D(\bk)}\right].
\label{eqC3v-chibartr-explicit-supp}
\end{equation}
Finally, using $H_{zxx}^{(s)}(\bk)=-\calQ_{xxz}^{(s)}(\bk)$ and $v_x^{(s)}(\bk)=\hbar^{-1}\bigl[k_x/m^*+s\,B_x(\bk)/\sqrt{D(\bk)}\bigr]$, the mixed-quadrupole coefficient is
\begin{equation}
\bar\chi_H=\beta^{(H)}_{zxxx}=e^3\tau\sum_{s=\pm}\int_{\Lambda}\calQ_{xxz}^{(s)}(\bk)\,v_x^{(s)}(\bk)\,f_0'(\eps_s).
\label{eqC3v-chibarH-explicit-supp}
\end{equation}
This is the explicit integral form of the mixed-quadrupole coefficient for the model. Given a magnetic coupling for the interband element $C_{\mu\nu}^z=-m_{\mu\nu}^z/\Delta_{\nu\mu}$, the numerical recipe is to compute
\begin{equation}
m^z_{\mu\nu}(\bk)=\frac{\ii e}{2\hbar}\epsilon_{zab}
\langle\partial_a u_\mu|\bigl[h(\bk)-\tfrac12(\eps_\mu+\eps_\nu)\bigr]|\partial_b u_\nu\rangle,
\label{eqinterband-moment-benchmark-supp}
\end{equation}
form $C^z_{s\bar s}=-m^z_{s\bar s}/\Delta_s$ and $R^x_{s\bar s}=A^x_{s\bar s}/\Delta_s$, and, since the $V$ sum is empty for a strict two-band model, evaluate
\begin{equation}
\begin{split}
\calQ^{(s)}_{xxz}=\mathrm{Re}\bigl[&2U^{s\bar s}_{xxz}[R,C]+2U^{s\bar s}_{xxz}[C,R]
-U^{\bar s s}_{xzx}[R,C]-U^{\bar s s}_{xzx}[C,R]\bigr],\\
H^{(s)}_{zxx}&=-\calQ^{(s)}_{xxz}.
\end{split}
\label{eqC3v-H-numerical-procedure-supp}
\end{equation}
The covariant derivative in $U$ is evaluated either analytically from the smooth eigenvectors or numerically by central differences after parallel transporting neighboring eigenvectors. An equivalent check is the finite-field extraction
\begin{equation}
H^{(s)}_{zxx}(\bk)=\lim_{E_x\to0}
\frac{m^{(s)}_z(\bk,E_x)-2m^{(s)}_z(\bk,0)+m^{(s)}_z(\bk,-E_x)}{2e^2E_x^2},
\label{eqH-finite-field-extraction-supp}
\end{equation}
with $m^{(s)}_z(\bk,E_x)$ evaluated on the field-corrected wave-packet state.

At zero temperature one has $f_0'(\eps_s)=-\delta(\eps_s)$, so Eq.~\eqref{eqC3v-chibarG-explicit-supp} reduces exactly to a Fermi-surface integral. Defining
$$
\mathcal F_s(\phi)\equiv \{\,k>0 \mid \eps_s(k,\phi)=0\,\},
$$
and the auxiliary integrand (not to be confused with the band-normalized metric $\mathcal G^\nu_{jk}$)
$$
\mathcal I(k,\phi)\equiv
\frac{\bigl[w_x(k,\phi)D(k,\phi)-2w(k,\phi)B_x(k,\phi)\bigr]
\bigl[D(k,\phi)\bigl(v^2+\lambda^2 w_x(k,\phi)^2\bigr)-B_x(k,\phi)^2\bigr]}
{D(k,\phi)^{9/2}},
$$
Eq.~\eqref{eqC3v-chibarG-explicit-supp} becomes
$$
\bar\chi_G^{T=0}
=
-\frac{e^4\tau\lambda v^2}{8\hbar^2(2\pi)^2}
\sum_{s=\pm}s
\int_0^{2\pi}\dd\phi
\sum_{\alpha\in\mathcal F_s(\phi)}
\frac{k_{s,\alpha}(\phi)\,
\mathcal I\!\bigl(k_{s,\alpha}(\phi),\phi\bigr)}
{\left|\partial_k\eps_s\!\bigl(k_{s,\alpha}(\phi),\phi\bigr)\right|},
$$
with
$$
\partial_k\eps_s(k,\phi)
=
\frac{k}{m^*}
+
s\,\frac{v^2+3\lambda^2k^4\cos^2 3\phi}
{\sqrt{v^2+\lambda^2k^4\cos^2 3\phi}}.
$$
For generic finite $\lambda$, the remaining angular integral is not elementary because $k_{s,\alpha}(\phi)$ is determined implicitly by a sextic equation. In the weak-warping regime $|\lambda|k_F^2\ll v$, Eq.~\eqref{eqC3v-chibarG-explicit-supp} simplifies to
$$
\bar\chi_G
=
-\frac{e^4\tau\lambda}{32\pi\hbar^2 v}
\sum_{s=\pm}s
\int_0^\infty \frac{\dd k}{k^2}\,
f_0'\!\bigl(\eps_s^{(0)}(k)\bigr)
+O(\lambda^3),
\qquad
\eps_s^{(0)}(k)=\frac{k^2}{2m^*}-\mu+s\,vk.
$$
At $T=0$ and $\mu>0$, the positive Fermi roots are
$$
k_{F,\pm}^{(0)}
=
\sqrt{m^*2v^2+2m^*\mu}\mp m^* v,
\qquad
\left.\partial_k\eps_s^{(0)}\right|_{k_{F,s}^{(0)}}
=
\frac{\sqrt{m^*2v^2+2m^*\mu}}{m^*},
$$
so that the $s=\pm$ contributions, evaluated at $k_{F,\pm}^{(0)}$, combine to give
\begin{equation}
\bar\chi_G^{T=0}
=
\frac{e^4\tau\lambda}{32\pi\hbar^2\mu^2}
+
O(\lambda^3).
\label{eqchibarG-T0-supp}
\end{equation}

By contrast, Eq.~\eqref{eqC3v-chibartr-explicit-supp} is not universal for the continuum model. At small $\lambda$,
$$
\frac{w(\mathbf k)}{\eps(\mathbf k)^2}
=
\frac{k_x^3-3k_xk_y^2}{v^2(k_x^2+k_y^2)}
+
O(\lambda^2),
$$
so that
$$
\partial_x^3\!\left[\frac{w(\mathbf k)}{\eps(\mathbf k)^2}\right]
=
\frac{24\,k_y^2\bigl(k_x^4-6k_x^2k_y^2+k_y^4\bigr)}
{v^2(k_x^2+k_y^2)^4}
+
O(\lambda^2),
$$
which is scale free in the continuum regime and behaves as $1/k^2$. The two-dimensional continuum integral in Eq.~\eqref{eqC3v-chibartr-explicit-supp} is therefore cutoff sensitive and cannot be reduced to a unique cutoff-independent number from Eq.~\eqref{eqC3v-benchmark-supp} alone. It is best written as
\begin{equation}
\bar\chi_{\mathrm{tr}}
=
\frac{e^4\tau^3\lambda}{\hbar^4}
\mathcal C_{\mathrm{tr}}(m^*,v,\mu,\Lambda,\text{regularization}),
\label{eqchibartr-nonuniv-supp}
\end{equation}
with $\mathcal C_{\mathrm{tr}}$ a nonuniversal dimensionless number fixed only after specifying the lattice completion or ultraviolet regularization of the continuum model.

The dimensionless model curves use the low-energy units $\hbar=e=v=m^*=1$, and the lattice figure additionally sets $a=1$. The vertical axis is $|\bar\chi_X|$ itself, not $|\bar\chi_X/\lambda|$, and the weak-warping factor $\lambda$ is retained explicitly, with the dashed guide in the metric plot given by $\bar\chi_G=\lambda/(32\pi\mu^2)$. In relaxation-time scans the band-geometry integral is fixed and only the explicit $\tau$ or $\tau^3$ prefactor is varied. 

Finally, Eq.~\eqref{eqC3v-chibarH-explicit-supp} reduces to the Fermi surface at $T=0$,
\begin{equation}
\bar\chi_H^{T=0}
=
-\frac{e^3\tau}{(2\pi)^2}
\sum_{s=\pm}
\int_0^{2\pi}\dd\phi
\sum_{\alpha\in\mathcal F_s(\phi)}
\frac{k_{s,\alpha}(\phi)\,\calQ_{xxz}^{(s)}\!\bigl(k_{s,\alpha}(\phi),\phi\bigr)\,v_x^{(s)}\!\bigl(k_{s,\alpha}(\phi),\phi\bigr)}{\left|\partial_k\eps_s\!\bigl(k_{s,\alpha}(\phi),\phi\bigr)\right|
}.
\label{eqchibarH-T0-supp}
\end{equation}
This is the most explicit form available within the continuum model. Thus only $\bar\chi_G$ has a universal weak-warping continuum coefficient. The coefficient $\bar\chi_{\mathrm{tr}}$ requires an ultraviolet regularization, and $\bar\chi_H$ requires a microscopic magnetic-coupling prescription or a lattice completion.

\section{Triangular-lattice regularization of the \texorpdfstring{$C_{3v}$}{C3v} model}
\label{sectriangular-regularization}
The continuum model in Eq.~\eqref{eqC3v-benchmark-supp} leaves the two issues
noted above, namely the magnetic-coupling input required by the
mixed-quadrupole contribution and the cutoff dependence of the transport
coefficient. Both are removed by a minimal periodic two-band completion on the
triangular Bravais lattice, in the same spirit as the hexagonal-warping
lattice completion of Ref.~\cite{Fu2009} and the symmetry-based
nonlinear-transport lattice models of Ref.~\cite{fang2024quantum}. We
therefore place one effective spin-$1/2$ doublet
$$
\Psi_{\mathbf R}=\bigl(c_{\mathbf R\uparrow},c_{\mathbf R\downarrow}\bigr)^{\mathsf T}
$$
on each lattice site, with the Pauli matrices acting in this pseudospin space.
This is the minimal time-reversal-symmetric realization of
Eq.~\eqref{eqC3v-benchmark-supp}.

The triangular Bravais lattice is generated by
\begin{equation}
\mathbf a_1=a(1,0),
\qquad
\mathbf a_2=a\left(\frac12,\frac{\sqrt3}{2}\right),
\label{eqtriangular-lattice-vectors-supp}
\end{equation}
and it is convenient to choose the three positive nearest-neighbor bonds as
\begin{equation}
\boldsymbol\delta_1=\mathbf a_1,
\qquad
\boldsymbol\delta_2=\mathbf a_2,
\qquad
\boldsymbol\delta_3=\mathbf a_2-\mathbf a_1=a\left(-\frac12,\frac{\sqrt3}{2}\right),
\label{eqtriangular-bonds-supp}
\end{equation}
with $\hat{\boldsymbol\delta}_n\equiv\boldsymbol\delta_n/a$. A convenient
tight-binding representative is
\begin{align}
H_\triangle&=H_0+H_R+H_w,
\label{eqtriangular-realspace-supp}\\
H_0&=\Delta_0\sum_{\mathbf R}\Psi_{\mathbf R}^\dagger\Psi_{\mathbf R}-t\sum_{\mathbf R,n}\Big(\Psi_{\mathbf R}^\dagger\Psi_{\mathbf R+\boldsymbol\delta_n}+\mathrm{h.c.}\Big),
\label{eqtriangular-H0-supp}\\
H_R&=\ii\sum_{\mathbf R,n}\Psi_{\mathbf R}^\dagger\Big[(\hat{\mathbf z}\times\hat{\boldsymbol\delta}_n)\cdot\boldsymbol\sigma\Big]\Big(t_1\Psi_{\mathbf R+\boldsymbol\delta_n}+t_2\Psi_{\mathbf R+2\boldsymbol\delta_n}\Big)+\mathrm{h.c.},
\label{eqtriangular-HR-supp}\\
H_w&=\ii t_3\sum_{\mathbf R}\Psi_{\mathbf R}^\dagger\sigma_z\Big(\Psi_{\mathbf R+2\boldsymbol\delta_1}-\Psi_{\mathbf R+2\boldsymbol\delta_2}+\Psi_{\mathbf R+2\boldsymbol\delta_3}\Big)+\mathrm{h.c.}.
\label{eqtriangular-Hw-supp}
\end{align}
Here $\Delta_0$ is an on-site energy, $t$ generates the parabolic scalar
dispersion, $t_1$ is the nearest-neighbor Rashba coupling, $t_2$ is an
auxiliary longer-range Rashba term that will be fixed by exact matching, and
$t_3$ is the inversion-breaking third-neighbor warping amplitude. In momentum
space,
\begin{equation}
H_\triangle=\sum_{\mathbf k}\Psi_{\mathbf k}^\dagger\,\mathcal H_\triangle(\mathbf k)\,\Psi_{\mathbf k},
\qquad
\mathcal H_\triangle(\mathbf k)=h_0(\mathbf k)\sigma_0+\mathbf d(\mathbf k)\cdot\boldsymbol\sigma,
\label{eqtriangular-bloch-supp}
\end{equation}
with $q_n\equiv\mathbf k\cdot\boldsymbol\delta_n$,
\begin{align}
h_0(\mathbf k)&=\Delta_0-2t\sum_{n=1}^3\cos q_n,
\label{eqtriangular-h0-supp}\\
\mathbf d_\parallel(\mathbf k)&=2\sum_{n=1}^3(\hat{\mathbf z}\times\hat{\boldsymbol\delta}_n)\Big[t_1\sin q_n+t_2\sin(2q_n)\Big],
\label{eqtriangular-dparallel-supp}\\
d_z(\mathbf k)&=-2t_3\Big[\sin(2q_1)-\sin(2q_2)+\sin(2q_3)\Big].
\label{eqtriangular-dz-supp}
\end{align}
With the standard time-reversal operator $\mathcal T=\ii\sigma_y K$, one has
$\mathcal T\mathcal H_\triangle(\mathbf k)\mathcal T^{-1}=\mathcal H_\triangle(-\mathbf k)$,
while inversion is broken by the odd pseudospin-dependent harmonics.

Expanding near $\Gamma$ gives
\begin{align}
h_0(\mathbf k)&=(\Delta_0-6t)+\frac{3ta^2}{2}(k_x^2+k_y^2)+\order{k^4},
\label{eqtriangular-h0-expand-supp}\\
\mathbf d_\parallel(\mathbf k)&=3a(t_1+2t_2)(k_x\hat{\mathbf y}-k_y\hat{\mathbf x})
-3a^3\left(\frac{t_1}{8}+t_2\right)(k_x^2+k_y^2)(k_x\hat{\mathbf y}-k_y\hat{\mathbf x})+\order{k^5},
\label{eqtriangular-dparallel-expand-supp}\\
d_z(\mathbf k)&=2a^3t_3\,(k_x^3-3k_xk_y^2)+\order{k^5}.
\label{eqtriangular-dz-expand-supp}
\end{align}
Equations~\eqref{eqtriangular-dparallel-expand-supp} and
\eqref{eqtriangular-dz-expand-supp} show that a nearest-neighbor Rashba term by
itself generates an unwanted isotropic cubic correction proportional to
$k^2(k_x\sigma_y-k_y\sigma_x)$. Exact reproduction of
Eq.~\eqref{eqC3v-benchmark-supp} through cubic order therefore requires
\begin{equation}
t_2=-\frac{t_1}{8}.
\label{eqtriangular-t2-match-supp}
\end{equation}
This constraint does not add a new low-energy parameter. It merely fixes the
auxiliary longer-range Rashba amplitude. Matching the remaining coefficients
to Eq.~\eqref{eqC3v-benchmark-supp} gives
\begin{equation}
t=\frac{1}{3m^*a^2},
\qquad
t_1=\frac{4v}{9a},
\qquad
t_2=-\frac{v}{18a},
\qquad
t_3=\frac{\lambda}{2a^3},
\qquad
\Delta_0=6t-\mu.
\label{eqtriangular-match-conditions-supp}
\end{equation}
The resulting Bloch Hamiltonian obeys
\begin{equation}
\mathcal H_\triangle(\mathbf k)=\left(\frac{k_x^2+k_y^2}{2m^*}-\mu\right)\sigma_0+v(k_x\sigma_y-k_y\sigma_x)+\lambda(k_x^3-3k_xk_y^2)\sigma_z+\cdots,
\label{eqtriangular-kp-match-supp}
\end{equation}
where the omitted terms begin at $\order{k^4}$ in the scalar channel and
$\order{k^5}$ in the Pauli-vector channel. Equation~\eqref{eqtriangular-kp-match-supp}
is therefore an exact $\Gamma$-point regularization of
Eq.~\eqref{eqC3v-benchmark-supp} to the order retained in the continuum model.

The two lattice bands are
$$
\eps_s(\mathbf k)=h_0(\mathbf k)+s\,\eps(\mathbf k),
\qquad
\eps(\mathbf k)=|\mathbf d(\mathbf k)|,
\qquad
P_s(\mathbf k)=\frac{1+s\,\hat{\mathbf d}(\mathbf k)\cdot\boldsymbol\sigma}{2},
\qquad s=\pm,
$$
with $\hat{\mathbf d}=\mathbf d/\eps$. Because $\mathcal H_\triangle(\mathbf k)$
is a finite $2\times2$ Bloch matrix on a compact Brillouin zone, the Berry
curvature, quantum metric, orbital moment, and interband matrix elements
entering Eqs.~\eqref{eqC3v-Q-formal-supp} and \eqref{eqC3v-RC-H-formal-supp}
are all fixed unambiguously. Standard two-band formulas give~\cite{xiao2010berry}
\begin{equation}
\Omega^z_s(\mathbf k)=-\frac{s}{2\eps(\mathbf k)^3}\,\mathbf d\cdot\Big(\partial_{k_x}\mathbf d\times\partial_{k_y}\mathbf d\Big),
\qquad
g^{(s)}_{ij}(\mathbf k)=\frac{1}{4}\,\partial_i\hat{\mathbf d}\cdot\partial_j\hat{\mathbf d},
\qquad
m^z_s(\mathbf k)=\frac{e}{\hbar}\bigl[\eps_s(\mathbf k)-h_0(\mathbf k)\bigr]\Omega^z_s(\mathbf k).
\label{eqtriangular-two-band-geometry-supp}
\end{equation}
The interband orbital matrix element $m_z^{\,s\bar s}(\mathbf k)$, and hence
$C_z^{\,s\bar s}$, $\calQ^{(s)}_{\alpha\beta\gamma}$, and $H_{zjk}^{(s)}$,
then follow directly from the same eigenvectors via
Eqs.~\eqref{eqinterband-moment-benchmark-supp} through
\eqref{eqC3v-H-numerical-procedure-supp}, so the previously formal
mixed-quadrupole contribution becomes a definite Brillouin-zone integral.

A practical numerical workflow is straightforward. The reciprocal lattice
vectors are
\begin{equation}
\mathbf b_1=\frac{2\pi}{a}\left(1,-\frac{1}{\sqrt3}\right),
\qquad
\mathbf b_2=\frac{2\pi}{a}\left(0,\frac{2}{\sqrt3}\right),
\label{eqtriangular-reciprocal-supp}
\end{equation}
and one may sample $\mathbf k=(n_1/N)\mathbf b_1+(n_2/N)\mathbf b_2$ on a
uniform $N\times N$ grid with $N$ between $200$ and $500$. Diagonalizing
$\mathcal H_\triangle(\mathbf k)$ yields $\eps_s(\mathbf k)$ and the
cell-periodic eigenvectors, from which one computes $\Omega^z_s(\mathbf k)$,
$m^z_s(\mathbf k)$, $g^{(s)}_{ij}(\mathbf k)$, and the interband matrix
elements entering $\calQ^{(s)}_{\alpha\beta\gamma}$. The three dc contributions
then follow from the compact Brillouin-zone versions of
Eqs.~\eqref{eqbetaH-integral-supp} through \eqref{eqbetatr-integral-supp}, or
from the finite-frequency ordered formulas in
Eq.~\eqref{eqbeta-low-general-supp}. In contrast to the continuum model, the
transport contribution is now cutoff independent and the weak-warping
approximation is optional rather than mandatory. This lattice regularization
turns the $C_{3v}$ continuum model into a fully computable toy model for
contribution-resolved numerical comparisons.

\begin{figure}[t]
\centering
\includegraphics[width=0.998\textwidth]{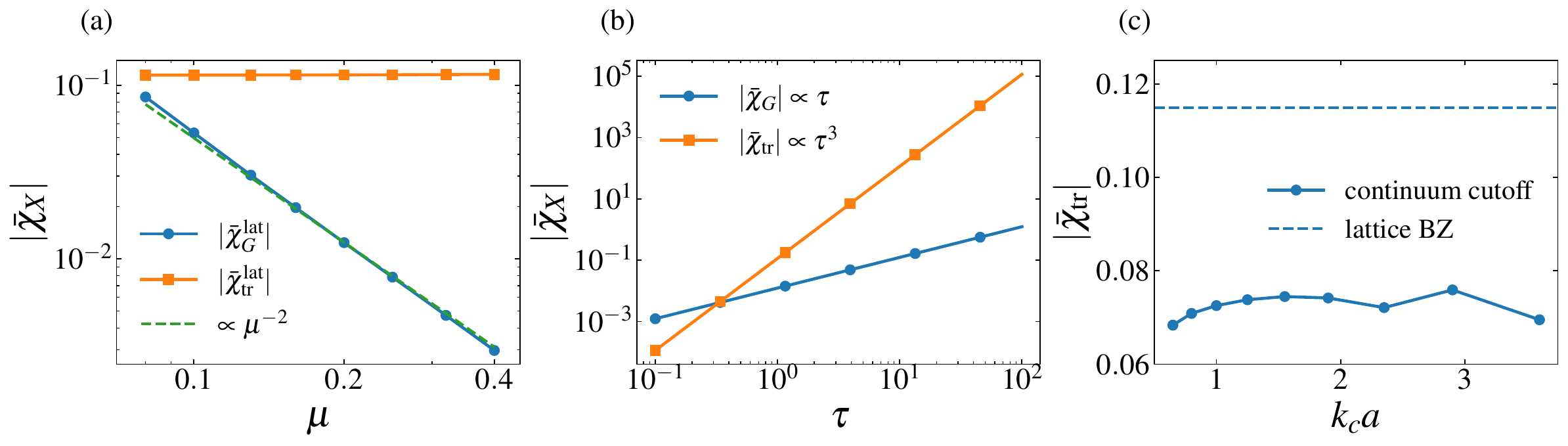}
\caption{
Triangular-lattice regularization of the $C_{3v}$ model. The plotted data use
$e=\hbar=a=v=m^*=1$, $\lambda=0.05$, temperature $T=0.008$, a midpoint
$241\times241$ Brillouin-zone grid, and finite-difference step $2\times10^{-3}$.
Panel (a) uses $\tau=1$ and $\mu=0.08,0.10,0.13,0.16,0.20,0.25,0.32,0.40$. The dashed guide is the weak-warping continuum prediction $\bar\chi_G=\lambda/(32\pi\mu^2)$. The computed values at $\mu=0.20$ are $|\bar\chi_G^{\rm lat}|=1.24\times10^{-2}$ and $|\bar\chi_{\rm tr}^{\rm lat}|=1.15\times10^{-1}$. Panel (b) fixes $\mu=0.20$ and varies $\tau$ from $10^{-1}$ to $10^2$, showing $|\bar\chi_G|\propto\tau$ and $|\bar\chi_{\rm tr}|\propto\tau^3$. Panel (c) compares the lattice value with a continuum transport integral using the soft cutoff $\exp[-(k/k_c)^2]$ on a $301\times301$ grid, for $k_c a=0.65,0.80,1.00,1.25,1.55,1.90,2.35,2.90,3.60$.
}
\label{fig2supp}
\end{figure}

Figure~\ref{fig2supp} validates the triangular-lattice completion. The metric
Brillouin-zone integral agrees with the weak-warping continuum prediction at
small $\mu$, confirming that the lattice Hamiltonian reproduces the continuum
$C_{3v}$ theory near $\Gamma$, and the relaxation-time scan reproduces the
$\tau$ and $\tau^3$ powers. Crucially, the lattice transport coefficient is
finite and cutoff independent, unlike its continuum counterpart. The
mixed-quadrupole contribution is not plotted, since it requires the separate
procedure of Eqs.~\eqref{eqinterband-moment-benchmark-supp} through
\eqref{eqC3v-H-numerical-procedure-supp}.

\section{Experimental signatures of cubic nonlinear magnetization}
\label{secexperiment}

This section translates the theoretical results into experimentally accessible observables. We focus on the angular fingerprint of the cubic tensor, THG-MOKE as the detection platform, separation of the three quantum-geometric contributions through frequency and relaxation-time scaling, and candidate material classes.

\subsection{Angular fingerprint and polarimetric detection}
\label{subsecangular}

The cleanest experimental handle on the cubic magnetization tensor is its characteristic
dependence on the azimuthal orientation $\phi$ of the driving electric field. For a
$C_{3v}$-symmetric sample with the mirror plane along $\hat x$ and an in-plane
field $\mathbf{E}_\omega = E_0(\cos\phi,\sin\phi,0)$, Eqs.~\eqref{eqC3v-2D-supp} and
\eqref{eqC3v-angle-supp} give the out-of-plane magnetization $M_z^{(3)} = \chi\,E_0^3\sin3\phi$,
or $\bar\chi\,E_0^3\cos3\phi$ in the rotated convention of the model
Hamiltonian~\eqref{eqC3v-benchmark-supp}. A full $\phi$ scan of $M_z$ at fixed $E_0$
constitutes the angular fingerprint measurement.

\subsection{THG-MOKE as an experimental platform}
\label{subsecmoke}

The cubic magnetization $M_i^{(3)}\propto E_j E_k E_l$ can be probed directly by third-harmonic magneto-optical Kerr spectroscopy~\cite{qian2026probing,morimoto2016semiclassical}. In a THG-MOKE measurement one drives the system at a single frequency $\omega$ and reads out the third-harmonic component of the magnetization at $\Omega=3\omega$. This output channel probes the frequency-resolved tensor $\beta^{\mathrm{phys}}_{ijkl}(3\omega,\omega,\omega,\omega)$ and measures a coherent cubic magnetization rather than a contribution-resolved observable. The dc expressions used in the main text are the leading quasi-static limit of this tensor. When $\omega\tau\gtrsim1$ or when interband resonances are approached, the ordered finite-frequency denominators in Eqs.~\eqref{eqfreq-H} through \eqref{eqfreq-tr} must be retained. In the single-$\tau$ regime the two geometric contributions $\beta^{(H)}$ and $\beta^{(G)}$ are both linear in $\tau$ and are therefore favored over the $\tau^3$ transport contribution at moderate disorder. Separating the mixed quadrupole from the metric contribution requires the gate and two-color frequency diagnostics described below. This configuration is the natural cubic counterpart of the SHG-MOKE methodology recently used by Qian \emph{et al.}~\cite{qian2026probing} to detect a second-order quantum-geometric magnetization in WTe$_2$. The same calibrated SMOKE setup can be retuned from $2\omega$ readout to $3\omega$ readout.

The measured Kerr rotation is proportional to the magneto-optical source at the emitted frequency, but the proportionality factor is sample-geometry dependent. A safe way to write the extraction is
\begin{equation}
\theta_K(3\omega)+\ii\eta_K(3\omega)=
\mathcal F(3\omega, n_s,d,\sigma_{\rm opt},\ldots)\,M_z^{(3)}(3\omega),
\label{eqkerr-formula}
\end{equation}
where $\mathcal F$ is obtained from the optical transfer matrix, including substrate index $n_s$, film thickness $d$, sheet or bulk optical conductivities, and boundary conditions. In the idealized normal-incidence bulk limit in Gaussian units, this reduces schematically to the familiar magneto-optical estimate~\cite{argyres1955theory,bennett1965faraday,zvezdin1997modern}
\begin{equation}
\theta_K+\ii\eta_K\simeq \frac{4\pi}{c}\frac{M_z^{(3)}}{n^2-1},
\label{eqkerr-bulk-limit-supp}
\end{equation}
with complex refractive index $n$. For a topological-insulator surface state, $M_z^{(3)}$ is a sheet magnetization rather than a bulk density. One may quote an effective bulk value by dividing by an optical thickness, but quantitative extraction should include the sheet response as a boundary condition in the transfer matrix. Substituting $M_z^{(3)}=\bar\chi E_0^3\cos3\phi$ gives a Kerr signal that can be read out by rotating the linear polarization of the pump laser.

\subsection{Frequency and relaxation-time scaling as contribution diagnostics}
\label{subsecscaling}

The three dc contributions carry distinct powers of the relaxation time $\tau$ and distinct
frequency denominators in the low-frequency tensor of Eq.~\eqref{eqbeta-low-TR-supp}.
These differences constitute independent diagnostics.

\paragraph*{$\tau$-scaling.}
From Eqs.~\eqref{eqbetaH-integral-supp} through \eqref{eqbetatr-integral-supp}
\begin{equation}
\beta^{(H)} \propto e^3\tau, \qquad
\beta^{(G)} \propto e^3\tau, \qquad
\beta^{(\mathrm{tr})} \propto e^3\tau^3.
\label{eqtau-scaling}
\end{equation}
At low disorder ($\tau$ large), the transport contribution ($\tau^3$) can dominate. At higher disorder ($\tau$ small), the mixed-quadrupole and metric contributions ($\tau^1$) are favored. Measuring the cubic magnetization as a function of sample purity, for example by varying temperature or controlled impurity doping, can therefore separate the contributions. The slope of $\ln|\bar\chi|$ versus $\ln\tau$ crosses over from $3$ in the clean transport-dominated limit to $1$ in the geometric-contribution-dominated limit. Side-jump, skew-scattering, multiband vertex corrections, and energy-dependent lifetimes can change the quantitative scaling in real samples. In particular, side-jump contributions are typically $\tau$-independent and could mimic an apparent geometric scaling in the moderate-disorder window, whereas skew-scattering produces $\tau$-dependent extrinsic pieces whose exponents depend on the dominant scattering mechanism. The clean separation in Eq.~\eqref{eqtau-scaling} should therefore be regarded as the reference structure of the intrinsic response, and is most reliably tested when combined with the cutoff-independent gate dependence $\bar\chi_G\propto\mu^{-2}$ derived below and with non-degenerate two-color frequency scans, both of which are insensitive to the lifetime in different ways. Analogous intrinsic-versus-extrinsic discussions in the nonlinear Hall context~\cite{lai2021third,liu2022berry,he2024third} confirm that this diagnostic is informative but not decisive on its own.

\paragraph*{Frequency denominators.}
From Eq.~\eqref{eqbeta-low-TR-supp}, the three surviving contributions at finite frequency
carry the Drude-like denominators
\begin{align}
\beta^{(H)} &\propto \frac{1}{1-\ii\omega_3\tau},
\label{eqfreq-H}\\
\beta^{(G)} &\propto \frac{1}{1-\ii\Omega\tau},
\label{eqfreq-G}\\
\beta^{(\mathrm{tr})} &\propto
\frac{1}{(1-\ii\Omega\tau)[1-\ii(\omega_2+\omega_3)\tau](1-\ii\omega_3\tau)}.
\label{eqfreq-tr}
\end{align}
The mixed-quadrupole contribution responds at the frequency of the last field insertion $\omega_3$, while the metric contribution responds at the total output frequency $\Omega$. These rolloffs refer to the ordered kernel at fixed $(j,\omega_1),(k,\omega_2),(l,\omega_3)$. The physical THG response is obtained by the symmetrization in Eq.~\eqref{eqbeta-phys-sym-supp}, which averages over which input leg carries which Cartesian index. For a degenerate single-color drive, $\omega_1=\omega_2=\omega_3=\omega$ and $\Omega=3\omega$. The resulting onset scales are therefore at $\omega$ for the mixed quadrupole, at $3\omega$ for the metric contribution, and at $\omega$, $2\omega$, and $3\omega$ for the transport contribution. The denominators should not be interpreted as suppressing the transport contribution in the strict $\omega\tau\ll1$ limit. There all denominators approach unity and the dc powers in Eq.~\eqref{eqtau-scaling} control the hierarchy. Their diagnostic value is in the rolloff regime $\omega\tau\gtrsim1$, where the transport contribution contains two additional Drude factors relative to the metric contribution and one additional factor relative to the mixed-quadrupole contribution. Non-degenerate two-color or three-color configurations give a substantially cleaner separation because $\omega_3$, $\omega_2+\omega_3$, and $\Omega$ can be tuned independently, allowing the per-contribution rolloffs to be measured one at a time rather than as a sum.

\paragraph*{Temperature dependence.}
At low temperatures $T\ll\mu$, the Fermi-surface integrals in
Eqs.~\eqref{eqbetaH-FS-supp} and \eqref{eqbetaG-FS-supp} are essentially
temperature-independent, and the only $T$-dependence enters through $\tau(T)$ via
electron-phonon scattering. In the two-band weak-warping limit the metric contribution reduces to Eq.~\eqref{eqchibarG-T0-supp}, $\bar\chi_G^{T=0}\propto\lambda/\mu^2$, so varying the chemical potential by gating or doping should produce a $\mu^{-2}$ dependence of the metric contribution to the cubic MOKE signal.

\subsection{Background discrimination and inverse-Faraday contributions}
\label{subsecbackgrounds}

Any THG-MOKE measurement of cubic orbital magnetization must control for several backgrounds, namely ordinary optical third-harmonic generation, inverse-Faraday effects, drive-induced Oersted fields, and bolometric heating. We collect here the discriminants that distinguish the intrinsic cubic magnetization from each. The angular fingerprint $\cos3\phi$ or $\sin3\phi$ is itself a strong filter, because most backgrounds are isotropic in $\phi$ on a $C_{3v}$ surface. The phase-sensitive Kerr response and the strict $E_0^3$ power law provide two further independent handles. The usual inverse-Faraday source is proportional to $\mathrm{Im}\,\mathbf E_\omega\times\mathbf E_\omega^*$ and is quadratic in the optical field. It can affect a third-harmonic measurement only through additional mixing or field-induced changes. It vanishes identically for a linearly polarized drive, while the intrinsic cubic magnetization studied here is finite for linear polarization. Comparing linear and circular polarizations is thus a direct discriminant. A $\cos3\phi$ angular pattern that persists for purely linear polarization and tracks the predicted $\mu^{-2}$ gating cannot be the usual inverse-Faraday response. Drive-induced Oersted fields are similarly polarization sensitive and additionally depend on sample geometry, while bolometric heating is broadband, slow, and lacks the $C_{3v}$ angular harmonic. A background that simultaneously reproduces the $C_{3v}$ angular harmonic with linear polarization, the cubic power law in $E_0$, the gate-tunable $\mu^{-2}$ trend of the metric contribution, and the predicted intrinsic $\tau$ scaling in a disorder scan is implausible.

\subsection{Order-of-magnitude estimate of the THG-MOKE signal}
\label{subsecestimate}

It is useful to give a rough estimate of the expected Kerr rotation in the most direct platform. For a Bi$_2$Te$_3$ surface state with hexagonal warping $\lambda\sim2.5\times10^{-28}$~eV~m$^3$, surface velocity $v\sim4\times10^5$~m\,s$^{-1}$, chemical potential $\mu\sim0.1$~eV measured from the Dirac point, and a transport lifetime $\tau\sim10^{-13}$~s, Eq.~\eqref{eqchibarG-T0-supp} gives $\bar\chi_G\simeq9\times10^{-30}$~A\,$(\mathrm{V}/\mathrm{m})^{-3}$. Equivalently, $M_z^{(3)}/E_0^3\sim10^{-18}\,\mu_B\,$nm$^{-2}\,$(V/cm)$^{-3}$. Under a THz drive at $E_0\sim10^4$~V\,cm$^{-1}$, this produces $M_z^{(3)}\sim10^{-6}\,\mu_B$\,nm$^{-2}$ at $3\omega$. Because the signal scales as $E_0^3$, a stronger pulsed THz drive with $E_0\sim10^5$~V\,cm$^{-1}$ would enhance the same estimate by $10^3$ and give $M_z^{(3)}\sim10^{-3}\,\mu_B$\,nm$^{-2}$. The corresponding Kerr scale is enhanced by the same factor, although this high-field estimate should be used only while heating, bolometric response, Oersted fields, and nonperturbative corrections remain controlled. Substituting the baseline $10^4$~V\,cm$^{-1}$ estimate into a transfer-matrix calculation with an optical surface thickness of order a nanometer gives a Kerr rotation well below a microradian. The absolute number is sensitive to band-structure detail, optical transfer factors, and surface quality. The gate dependence of the metric channel, $\mu^{-2}$, and the polarization comparison against inverse Faraday are more robust than this absolute estimate.

\subsection{Candidate material platforms}
\label{subsecmaterials}

The most direct platforms are systems in which the angular fingerprint, gate tuning, and disorder or frequency scans are all accessible. $C_{3v}$ topological-insulator surfaces such as Bi$_2$Se$_3$ and Bi$_2$Te$_3$ realize the hexagonally warped Dirac structure of Eq.~\eqref{eqC3v-benchmark-supp}~\cite{Fu2009,xiao2010berry}. The $\cos3\phi$ fingerprint and the gate-controlled $\mu^{-2}$ metric test are therefore especially transparent. Zincblende or tetrahedral semiconductors provide cubic-leading axial responses from Dresselhaus spin-orbit coupling. Orthorhombic $T_d$ semimetals such as MoTe$_2$ and WTe$_2$ and moir\'e systems with $C_3$ or $C_{3v}$ symmetry are useful broader platforms because they already support nonlinear magneto-optical protocols and tunable band geometry, although lower-symmetry quadratic channels may coexist with the cubic signal.

\subsection{Summary of experimental predictions}
\label{subsecexp-summary}

Table~\ref{tabexp-predictions} summarizes the main experimental predictions.

\begin{table}[H]
\caption{Summary of experimental predictions for cubic nonlinear magnetization.
$\tau$ is the scattering time, $\mu$ the chemical potential, and $\phi$ the azimuthal
angle of the driving field.}
\label{tabexp-predictions}
\centering
\small
\begin{tabular}{llll}
Observable & Prediction & Diagnostic & Platform \\
\hline
Angular scan of $M_z$ & \makecell[l]{$\sin3\phi$ or $\cos3\phi$\\ (related by $30^\circ$ rotation)} & \makecell[l]{Identifies $C_{3v}$\\ symmetry class} & \makecell[l]{TI surface,\\ TMD} \\
Linear vs.\ circular polarization & \makecell[l]{Cubic $M_z$ finite for linear\\ usual inverse Faraday vanishes} & \makecell[l]{Rejects\\ inverse-Faraday background} & \makecell[l]{TI surface\\ TMD} \\
$\tau$-dependence (intrinsic) & \makecell[l]{$|\bar\chi|\propto\tau$ (geometric),\\ $|\bar\chi|\propto\tau^3$ (transport)} & \makecell[l]{Separates mixed-quadrupole/metric\\ from transport,\\ extrinsic side-jump/skew\\ can modify quantitative scaling} & \makecell[l]{Tunable\\ disorder} \\
Gate-voltage scan & $|\bar\chi_G|\propto\mu^{-2}$ & \makecell[l]{Tests metric contribution\\ formula Eq.~\eqref{eqchibarG-T0-supp},\\ cutoff-independent} & \makecell[l]{TI surface\\ gated} \\
Two-color frequency sweep & \makecell[l]{$\beta^{(H)}$ Drude rolloff at $\omega_3\tau\sim 1$,\\ $\beta^{(G)}$ rolloff at $\Omega\tau\sim 1$} & \makecell[l]{Resolves contribution\\ relaxation scales (only\\ non-degenerately)} & \makecell[l]{Two-color\\ THz} \\
THG MOKE & \makecell[l]{Coherent cubic\\ magnetization response} & \makecell[l]{Cubic ($\chi^{(3)}$) counterpart of the quadratic\\ $\chi^{(2)}$ Christoffel magnetization of Refs.~\cite{qiang2026quantum,qian2026probing}} & \makecell[l]{Noncentrosymmetric\\ metals} \\
Temperature & \makecell[l]{Tracks contribution Drude factors,\\ mixed-quadrupole/metric $\sim\tau(T)$, \\transport $\sim\tau^3(T)$} & \makecell[l]{Maps $e$-ph scattering\\ to response} & \makecell[l]{Low-T\\ spectroscopy} \\
\end{tabular}
\end{table}

\end{document}